\documentclass[
reprint,
superscriptaddress,
showpacs,
preprintnumbers,
nofootinbib,
nobibnotes,
amsmath,
amssymb, 
aps,
prd,
floatfix
]{revtex4-1}

\usepackage{booktabs}
\usepackage[utf8]{inputenc}
\usepackage[normalem]{ulem}
\usepackage{graphicx}
\usepackage{dcolumn}
\usepackage[colorlinks=true,allcolors=brown]{hyperref}
\usepackage{url}
\usepackage{enumerate}
\usepackage{bm}
\usepackage[dvipsnames]{xcolor}
\usepackage{mathrsfs}
\usepackage{amsmath}
\usepackage{cancel}
\usepackage{bbold}
 \usepackage{mathrsfs}
\usepackage{braket}
\usepackage{physics}
\usepackage{multirow}
\usepackage[capitalize]{cleveref}
\usepackage{xspace}
\usepackage{comment}
\usepackage{placeins}
\usepackage{fontawesome} 

\definecolor{palatinate}{rgb}{0.494, 0.192, 0.482}
\definecolor{blue-violet}{rgb}{0.33, 0.17, 0.89}
\renewcommand{\phi}{\varphi}

\def\absQ{Q}
\def\e{ \mathrm{e}}
\def\iu{ \mathrm{i}}

\begin{document}
\preprint{CALT-TH/2025-029~~~~CERN-TH-2025-178}
\title{Long-lived axion-like particles from electromagnetic cascades} 

\author{Samuel Patrone}
\email{spatrone@caltech.edu}
\affiliation{Walter Burke Institute for Theoretical Physics, California Institute of Technology, Pasadena, CA 91125, USA}
\author{Nikita Blinov}
\email{nblinov@yorku.ca}
\affiliation{Department of Physics and Astronomy, York University, Toronto, Ontario, M3J 1P3, Canada}
\author{Ryan Plestid}
\email{ryan.plestid@cern.ch}
\affiliation{Walter Burke Institute for Theoretical Physics, California Institute of Technology, Pasadena, CA 91125, USA}
\affiliation{Theoretical Physics Department, CERN, 1 Esplanade des Particules, CH-1211 Geneva 23, Switzerland}

\begin{abstract}
    We study axion-like particles (ALPs) in beam dump experiments, focusing on the Search for Hidden Particles (SHiP, at CERN) experiment and the Beam Dump eXperiment (BDX, at JLab). Many existing projections for sensitivity to ALPs in beam dump experiments have focused on production from either the primary proton/electron beam, or - in the case of SHiP - the secondary (high-energy) photons produced by neutral meson decays (e.g.,~$\pi^0\rightarrow\gamma \gamma$). In this work, we study the subsequent production of axions from the full electromagnetic shower in the target, finding order-of-magnitude enhancements in the visible decay yields across a wide range of axion masses. 
    We update SHiP's sensitivity curve and provide new projections for BDX. Both experiments will be able to reach currently unexplored regions of ALP parameter space.  
\end{abstract}

\maketitle

\section{Introduction \label{sec:Introduction} }
One of the simplest mechanisms that generate light, feebly interacting particles involves the spontaneous breaking of global symmetries at a high scale. The pseudo-Nambu-Goldstone bosons that arise are naturally light, with their masses protected by a shift symmetry.  Being derivatively coupled, their interactions are suppressed by a large decay constant $\sim |\bm{p}_a|/f_a$. These types of fields are often referred to as ``axion-like particles'' (or ALPs), and occur ubiquitously in many extensions of the Standard Model. In what follows, we will use ``axion'' and ``ALP'' interchangeably but do not restrict ourselves to models that solve the strong-CP problem~\cite{Wilczek:1977pj,Weinberg:1977ma,Kim:1979if,Shifman:1979if,Dine:1981rt,Zhitnitsky:1980tq}.

Feebly interacting particles, such as ALPs, can be discovered using beam-dump accelerator facilities~\cite{Beacham:2019nyx,Batell:2022dpx,Ilten:2022lfq}. The search strategy involves a high-energy (typically electron or proton) beam incident on a thick target. A detector is placed sufficiently far from the target so that beam-related backgrounds are minimized, and one searches for the scattering or decay of ALPs, or other particles, produced in the beam dump. 

\begin{figure}[t]
    \centering
    \includegraphics[width=\linewidth]{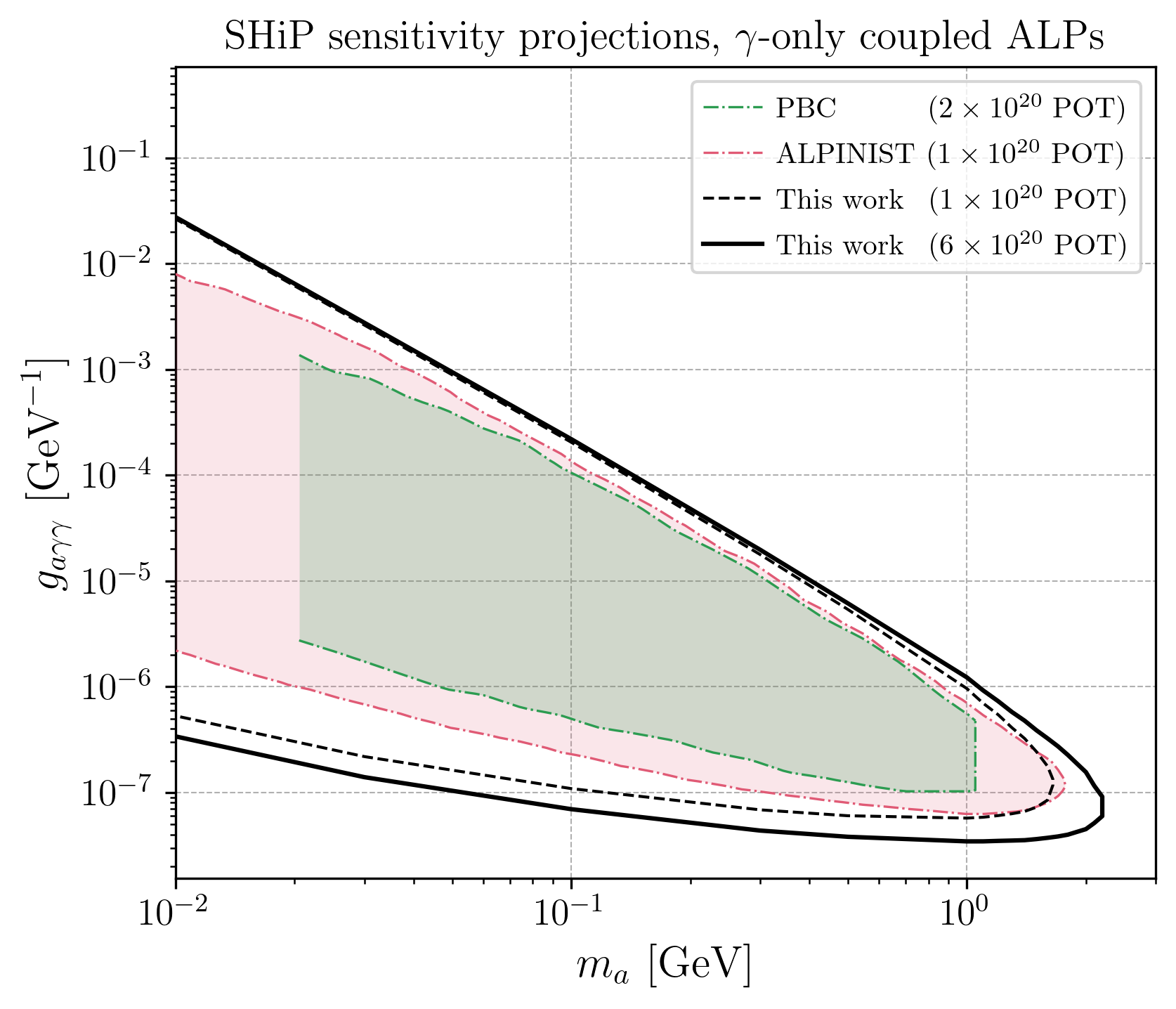}
    \caption{Comparison to past literature -- PBC~\cite{Beacham:2019nyx} (green shaded region), ALPINIST~\cite{Jerhot:2022chi} (red shaded region) -- of our projections for the sensitivity of SHiP (solid black line) to axion-photon couplings, with updated geometry and nominal protons-on-target (POT)~\cite{Aberle:2839677, Albanese:2878604}. For reference, we also plot our sensitivity projections with a lower POT, used before in the literature (dashed black line).  We defined $g_{a\gamma\gamma}\equiv f^{-1}c_{a\gamma\gamma}\alpha/\pi$, and we set $c_{aee}=0$. Accounting for the full electromagnetic cascade substantially enhances the reach of SHiP at lower couplings and lighter masses. See \cref{fig:ship_ecuts_sens} for an illustration of how this result depends on the energy threshold used at SHiP. 
    \label{fig:comparison-plot}}
\end{figure}

The purpose of this paper is to incorporate often neglected production mechanisms, in particular electromagnetic cascades, that enhance (see \cref{fig:comparison-plot})  the sensitivity to visibly decaying ALPs.  While these effects have been previously considered\footnote{Past literature often uses a one-dimensional ``forward-only approximation'' that overestimates geometric acceptances \cite{Blinov:2024pza}. } in the context of neutrino experiments (also proton beam dumps) \cite{Brdar:2020dpr,CCM:2021jmk,Capozzi:2023ffu} and the SLAC-E137 experiment \cite{Tsai:1986tx,Bjorken:1988as}, they are often neglected in the analysis of other beam dumps. Our goal is to clarify when and why enhancements from electromagnetic cascades are substantial, and to quantify their impact using two representative examples: SHiP (a proton beam dump) and BDX (an electron beam dump).

A common benchmark for comparison in the ALP literature (see e.g.,~\cite{Beacham:2019nyx}) is an ALP coupled to photons, 
\begin{equation}
\label{eq:lagrangian}
    -\mathcal{L} \supset \frac{1}{2} m_a^2 a^2  -  \frac{\alpha}{4\pi}\frac{c_{a\gamma \gamma}}{f} aF\tilde{F} ~ - \frac{c_{aee}}{f} (\partial_\mu a)\bar{e}\gamma^\mu \gamma_5 e ~, 
\end{equation}
where $a$ is the ALP field, $m_a$ its mass, $c_{a\gamma\gamma}$ ($c_{aee}$) its coupling to photons (electrons), $F$ is the electromagnetic field strength, and $\tilde{F}$ its dual. We have included an electron coupling, $c_{aee}$, which is naturally generated by loops involving photons or may appear at tree-level in an ultraviolet (UV) completion. Sometimes the coupling $g_{a\gamma\gamma}$ is used in the literature; following the conventions of e.g., \cite{Dobrich:2019dxc}, this is related to our notation via $g_{a\gamma\gamma} = c_{a\gamma\gamma} (\alpha/\pi)f^{-1}$.  As is well appreciated in the literature, high-intensity beam dump experiments can discover ALPs for masses $m_a\lesssim 2~{\rm GeV}$. 
  
The Search for Hidden Particles (SHiP) experiment is an ambitious next-generation beam dump facility~\cite{SHiP:2015vad,Aberle:2839677,Albanese:2878604}. Existing studies of SHiP \cite{Dobrich:2019dxc} have found that the dominant production mechanism is Primakoff scattering on nuclei. However, only virtual photons from the primary proton beam, and real photons from first-generation meson decays, have been included in flux estimates for SHiP. As we discuss in detail below, each of these real photons produces $10^2-10^3$ further secondary photons, electrons, and positrons while showering in the beam stop. Furthermore, the cascade particles produce lower-energy, and therefore shorter-lived, ALPs, increasing predicted yields at small masses and couplings. We find that the inclusion of electromagnetic cascades can enhance coupling sensitivity by more than an order of magnitude (corresponding to $\sim 10^4$ in the event rate). 

This enhancement is a consequence of the infrared (IR) sensitivity of the signal event rate. At small couplings, where the decay length is long, the detected event rate scales as
\begin{equation}
    R_{\text{det}} \propto \int \dd \omega_a ~  \dv{R_a}{\omega_a}  \frac{m_a}{\omega_a} ~, 
\end{equation}
where $m_a/\omega_a$ accounts for the time dilation in the axion's decay length. 
The rate of axion production $\dd R_a / \dd\omega_a$ inherits the IR-dominated spectrum of bremsstrahlung photons, being cut off at some $\omega_{\rm min} \sim m_a$. The event rate in the long-lifetime limit is therefore dominated by the lowest energy flux of axions from the beam stop; this will be produced by the electromagnetic cascade. 

The Beam-Dump eXperiment (BDX) at Jefferson Lab~\cite{BDX:2016akw} is an electron beam dump experiment that will search for the production of light dark matter, and its elastic and inelastic scattering in a downstream detector, reaching the limit of the irreducible neutrino  background. This setup will also be sensitive to a multitude of visibly-decaying dark sector states.
The incident electron beam will initiate an electromagnetic cascade in the thick target in which numerous secondary particles can produce ALPs. 
To our knowledge, a comprehensive study of the BDX setup for axion-like particles has not been undertaken (although see \cite{Chakrabarty:2019kdd} for an analysis with some of the relevant production mechanisms). 

In what follows we briefly review the effective theory of an ALP coupled to photons and electrons in \cref{sec:alp_eft}. We define our conventions and discuss renormalization group (RG) running below the weak scale, following \cite{Bauer:2020jbp}. In \cref{sec:production} we discuss the production mechanisms inside a thick target. We summarize our numerical methods in \cref{sec:methods}. In \cref{sec:experiments} we turn to the upcoming SHiP (400 GeV proton beam dump) experiment -- which produces a large photon flux, primarily via $\pi^0$ production, followed by $\pi^0\rightarrow \gamma\gamma$ -- and the BDX experiment at Jefferson Lab (10.6 GeV electron beam dump). We study the extent to which electromagnetic cascades enhance sensitivity to long-lived ALPs. We find a substantial (orders of magnitude) increase in the number of axions that are produced and decay within the experiments' decay pipe. Finally, we summarize our findings and conclude with an outlook towards interesting potential future directions in \cref{sec:conclusion}. 

\section{Axion couplings at low scales}
\label{sec:alp_eft}
We define two benchmark scenarios to illustrate phenomenology dominated by photons, and electrons, respectively. We fix the photon and electron couplings at $\mu_w=80 ~{\rm GeV}$ (just below the $W$-boson mass). This scale is low enough to talk sensibly about photons without considering $\gamma Z$ mixing, but high enough to ensure that realistic loop-induced couplings to electrons and photons are generated by renormalization group (RG) flow. We relate the couplings at $\mu_w$ to the coupling at $\mu_a=m_a$ using the RG flow described in~\cite{Bauer:2020jbp}.
Concretely, we take:
\begin{enumerate}[i)]
    \item {\it $\gamma$-dominated:}    $c_{a\gamma\gamma}(\mu_w)=1$, $c_{aee}(\mu_w)=0$~.
    \item {\it $e$-dominated:}  $c_{a\gamma\gamma}(\mu_w)=0$, $c_{aee}(\mu_w)=1$~.
\end{enumerate}
Although $c_{a\gamma\gamma}$ does not flow under RG, in phenomenological applications the replacement of $c_{a\gamma\gamma} \rightarrow \tilde{c}_{a\gamma \gamma}(\mu)$ captures effects generated by loops of light particles (such as the electron). The tilded coupling is defined \cite{Bauer:2020jbp} as:
\begin{equation}
   \tilde{c}_{a\gamma\gamma}(\mu) = c_{a\gamma \gamma} +  \sum_f N_c Q_f^2 c_{aff}(\mu) \Theta(\mu-m_f)
\end{equation}
with $f$ running over all fermions in the Standard Model, $Q_f$ the fermion electric charge, $N_c$ its number of colors, and $m_f$ its mass. The coupling $\tilde{c}_{a\gamma\gamma}$ naturally appears in the RG equations.  

In both benchmarks, at the scale of the axion mass both couplings are present, i.e., $c_{aee}(\mu_a)\neq 0$ and $\tilde{c}_{a\gamma\gamma}(\mu_a)\neq0$. This avoids pathologies where e.g., the computed axion bremsstrahlung rate vanishes identically. 

The relevant solutions of the RG flow are given by \cite{Bauer:2020jbp}, 
\begin{align}
    \label{caee-flow}
    c_{aee}(\mu_L) &= c_{aee}(\mu_H) - \frac{3 \tilde{c}_{a\gamma\gamma}(\mu_H)}{\beta_0^{\rm QED}} \frac{\alpha(\mu_L) - \alpha(\mu_H)}{\pi} ~\\
    &\simeq c_{aee}(\mu_H) + \frac{3 \alpha(\mu_H)}{4\pi} \log(\frac{\mu_H^2}{\mu_L^2}) \tilde{c}_{a\gamma\gamma}(\mu_H)~, \nonumber
\end{align}
where $\mu_H$ is the ``high scale'' (where initial conditions are set) and $\mu_L$ the ``low scale''. In the second line, we have expanded to $O(\alpha_H^2)$. For our numerical inputs, the RG flow of $\alpha(\mu)$ is solved with the one-loop QED beta function, with the relevant fermions integrated out at $\mu=m_f$. In practice, as can be readily seen from \cref{caee-flow}, the end result is not sensitive to these details since the $\beta_0^{\rm QED}$ does not appear at $O(\alpha^2)$. 

Next, consider the effective photon coupling $\tilde{c}_{a\gamma\gamma}$. We approximate:
\begin{equation}    
       \tilde{c}_{a\gamma\gamma} \approx c_{a\gamma \gamma} +   c_{aee}~,
    \label{eq:cagg_rg_val}
\end{equation}
where we neglect the running of $c_{aee} \simeq c_{aee}(\mu_w)$. This approximation is reliable for our purpose, because the two benchmarks we consider are dominated by either $c_{a\gamma\gamma} \gg c_{aff}$ or by $c_{aee} \gg c_{aff}$ for $f\neq e$. We have dropped the Heaviside function $\Theta(\mu-m_e)$ since we will always take $\mu = m_a$ and $m_a > m_e$. For the electron-dominated benchmark, our analysis matches with the conventions of~\cite{Eberhart:2025lyu}. For the photon-dominated benchmark, we have solved for the RG flow of $c_{aee}$. At a renormalization scale of $\mu_c=m_c$, we find $c_{aee}(\mu_c) = 3.6\times 10^{-5} c_{a \gamma \gamma}$.

Before proceeding, let us comment on how small hadronic couplings must be for electromagnetic production to dominate. This can be inferred from Fig.~9 of~\cite{Blinov:2021say}. In that work, an $O(1)$ coupling to gluons ($\frac{\alpha_s}{4\pi} c_{agg} G\tilde{G}$) was assumed. This induces a photon coupling of order $c_{a\gamma\gamma} \sim c_{agg}$, and hadronic production (from\footnote{In~\cite{Blinov:2021say}, only hadronic production that properly implements chiral symmetry \cite{Bauer:2021wjo} was included. Other production modes, such as the production of axions from the primary proton beam, were not included.} $\eta \rightarrow \pi \pi a$) was found to dominate in the first interaction length of the beam dump by $\sim 10^6$.

Any scenario that involves couplings to the electroweak or leptonic sector will only ``feed down'' into quarks at one-loop, producing couplings of  $c_{aqq} \sim (\alpha/4\pi)^2 \beta_0^{\rm QED} c_{a\gamma \gamma} $ and $c_{agg} \sim (\alpha/4\pi)^2 \beta_0^{\rm QED}  c_{a\gamma\gamma}$.
Since production rates scale as the square of the coupling, this naturally provides a $10^{-12}$ suppression for which electromagnetic modes easily dominate over hadronic production. 

Furthermore, in what follows, we find large enhancements to the cascade production (ranging from factors of $10^2-10^5$ depending on the axion mass). This is not sufficient to overcome hadronic production when $c_{agg} \sim c_{a\gamma\gamma}$; however, it can be enough for scenarios where, e.g., $c_{agg} \sim 10^{-1} \times c_{a\gamma\gamma}$. This may occur if the ratio of electromagnetic and QCD anomaly coefficients of the UV model, $E/N$,  is larger than the typical benchmark of $8/3\approx2.6$ \cite{Alonso-Alvarez:2018irt,Plakkot:2021xyx}, or via a modest kinetic mixing with some hidden sector \cite{Daido:2018dmu} among many other possible mechanisms \cite{Farina:2016tgd,Agrawal:2017cmd,DiLuzio:2020wdo}. A detailed study of the interplay between the shower-enhanced electromagnetic production versus hadronic production modes in specific UV completions warrants further study, but lies beyond the scope of this paper.

\section{Axion production mechanisms}
\label{sec:production} 
In an electron or proton beam dump, ALPs can originate via multiple production mechanisms.  As mentioned above, depending on the UV completion under consideration, either hadronic or electromagnetic production can dominate the phenomenology. Since the electromagnetic cascade is our focus, we do not discuss hadronic production modes further.  

Previous projections for the SHiP experiment have neglected electromagnetic shower-initiated production in the target, focusing only on the first interaction length.  In recent work~\cite{Blinov:2024pza,Zhou:2024aeu}, two of us have developed a framework for the systematic inclusion of shower-induced production of light long-lived particles, implemented in the public code \texttt{PETITE}. A key advantage of this approach is its computational efficiency, thanks to the extensive usage of pre-computed \texttt{VEGAS} integrators. 
For this work, we developed \href{https://github.com/spatrone/ALPETITE}{\faGithub~\texttt{ALPETITE}} \cite{github}, a dedicated tool that extends the \texttt{PETITE} framework to provide a comprehensive calculation of axion flux from electromagnetic cascades. Our approach is twofold. First, for production channels that have a direct analogue between dark photon and axion models (e.g., Bremsstrahlung), \texttt{ALPETITE} re-weights the simulated dark photon events generated by \texttt{PETITE} to derive the corresponding axion flux. Second, for processes unique to axion physics that lack a vector-particle equivalent (such as Primakoff production), we implemented and integrated dedicated Monte Carlo samplers. These samplers are not run in isolation; instead, they take the SM particle shower information generated by \texttt{PETITE} as their direct input. This hybrid strategy allows \texttt{ALPETITE} to combine the speed of re-weighting with the completeness of a full simulation, ensuring an accurate and efficient result.

We now describe the relevant electromagnetic processes that dominate the ALP production rate in an electromagnetic cascade. 
\paragraph{\bf Primakoff upscattering:} 
Photons can scatter on nuclei to produce ALPs of mass $m_a$ via the $c_{a\gamma\gamma}$ coupling, 
\begin{equation}
   \gamma  Z \rightarrow  a Z~,
\end{equation}
where $Z$ denotes a nucleus with charge $Z$. Treating the nucleus as a static charge density $\rho(\vb{x})$ that sources a Coulomb field, the differential cross-section is
\begin{equation}
\begin{split}
\label{eq:dsigma_primakoff}
    \dv{\sigma}{Q^2} =& \qty(\frac{1}{f})^2\frac{Z^2 \alpha^3 c^2_{a\gamma\gamma}}{32 \pi^2 Q^4\omega^2}\times \\ &\left(2Q^2(2\omega^2-m_a^2)- m_a^4-Q^4\right)|F_A(Q^2)|^2~,
\end{split}
\end{equation}
where 
\begin{equation}
    F_A(Q^2)=\int {\dd}^3x \; \e^{\iu \vb{q}\cdot \vb{x}}\rho(\vb{x})~,
\end{equation}
and $\omega$ is the energy of the incoming photon.
Because the energy transfer to the nucleus, $\Delta \omega\simeq \vb{q}^2/2M_A$, is negligible, the three-momentum and four-momentum transfer are nearly identical
\begin{equation}
    Q^2=-q^2\simeq \vb{q}^2~.
\end{equation}

Axions may also be produced via incoherent upscattering. This corresponds to inelastic final states for the nuclear system (e.g., nucleon knockout). For consistency with the treatment of dark vector production, we incorporate incoherent upscattering using a replacement rule for the form factor, 
\begin{equation}
    |F_A(\absQ^2)|^2 \rightarrow G_{2,\rm el}(\absQ^2) + G_{2,\rm inel}(\absQ^2)/Z~, 
\end{equation}
where the two contributions are given by~\cite{Bjorken:2009mm}
\begin{align}
    G_{2,\rm el}(\absQ^2)  &= \qty(\frac{a^2 \absQ^2}{1+ a^2\absQ^2})^2\qty(\frac{1}{1+\absQ^2/d^2})^2~, \label{eq:G_el}\\
    G_{2,\rm inel} (\absQ^2) &= \qty(\frac{a^{\prime2} \absQ^2}{1+ a^{\prime2}\absQ^2})^2 \qty(\frac{1+ \frac{\absQ^2}{4 m_p^2}(\mu_p^2-1)}{(1+ \absQ^2/\Lambda_H^2)^4})~.
    \label{eq:G_inel}
\end{align}
The atomic parameters are given by 
\begin{equation}
    a=\frac{111}{Z^{1/3}m_e}  \qq{and}   a'= \frac{773}{Z^{2/3}m_e} ~,
    \label{eq:atomic_para}
\end{equation}
while the hadronic parameters are
\begin{equation}
\Lambda_H=0.84~{\rm GeV}  \qq{and} d=0.405\, A^{-1/3}\, \mathrm{GeV}~.
\end{equation}
The proton's magnetic moment is $\mu_p=2.79$, and $m_p$ is the mass of the proton. 
It should be noted that the second bracketed term in \cref{eq:G_inel} is {\it not} squared,
see Footnote 4 of ~\cite{Celentano:2020vtu}. 

In past studies of SHiP, only the coherent Primakoff channel has been included. It is commonplace in dark photon searches to also include the inelastic production discussed above (see e.g.,~\cite{Bjorken:2009mm}, or Appendix B of either~\cite{Celentano:2020vtu} or \cite{Blinov:2024pza}).  This production channel can become important at larger axion masses and we therefore include it in our analysis. In \cref{app:inelastic_gain}, we show the (modest) gain in downstream flux from the inelastic contribution for SHiP with and without electromagnetic showers.

In addition to production from real photon beams, ALPs can also be produced when an electron beam impinges on a target via photon fusion, 
\begin{equation}
    e^- Z \xrightarrow{\gamma\gamma} a\,e^- Z 
\end{equation}
 This process is calculated using the Equivalent Photon Approximation (EPA), also known as the Weizsäcker-Williams method~\cite{vonWeizsacker:1934nji,Williams:1934ad}. In this framework, the total cross-section is obtained by convolving the on-shell Primakoff cross-section with the equivalent photon spectrum, expressed as a function of the photon energy fraction $x = \omega/E_e$:
\begin{equation}
    \label{eq:epa_primakoff_x}
    \sigma(eZ \to e Z a) = \int_{x_{\rm min}}^{1} \dd x \, \frac{\dd N_\gamma}{\dd x} \sigma_{\gamma Z \to a Z}(x E_e)~,
\end{equation}
where $E_e$ is the incident electron energy and $x_{\rm min}$ is an arbitrary IR cutoff. The ``doubly-logarithmic'' version of the EPA flux is given by~\cite{Budnev:1975poe}
\begin{equation}
\label{eq:epa_spectrum_x}
\frac{\dd N_\gamma}{\dd x} = \frac{\alpha}{\pi} \left[\frac{1+(1-x)^2}{x} \right]\log \frac{Q^2_{\rm max}}{Q^2_{\rm min}}~.
\end{equation}
Here, $Q^2_{\rm min} = m_e^2 x^2/(1-x)$ is the kinematically allowed minimum photon virtuality, and $Q^2_{\rm max}$ is a cutoff scale related to the target's structure. The photon flux is peaked at low $x$, providing a direct mechanism for ALP production. This channel is particularly important for fixed-target experiments utilizing primary electron beams, e.g.~BDX, providing a direct mechanism for ALP production that competes and -- as we will see in the following -- exceeds the primary electron bremsstrahlung.

\paragraph{\bf Axion bremsstrahlung:} Given a non-zero coupling to electrons, $c_{aee}$, one may also consider axion bremsstrahlung, 
\begin{equation}
    e^- Z \rightarrow a\,e^- Z ~. 
\end{equation}
This process has a matrix element structure that is similar to dark photon, $A_\mu'$, bremsstrahlung. It is instructive to study the ratio between the square matrix elements for the two processes. In the static limit (i.e., the nucleus is taken as infinitely heavy), both processes are allowed for $E_e > m$, where $m$ is either the vector or axion mass. In the ultra-relativistic limit ($E_e\gg m_e$), we find that
\begin{align}
    \label{eq:R-matrix-elements}
    \frac{\langle |\mathcal{M}_{a} |^2 \rangle}{\langle |\mathcal{M}_{V} |^2 \rangle} =  \frac{1}{2}\left(\frac{\epsilon_a}{\epsilon_V}\right)^2+\mathcal{O}\left(\frac{Q^2}{m^2}\right)
\end{align}
where $\epsilon_a\equiv2m_ec_{aee}/ef$, and the interaction lagrangian for a kinetically mixed dark photon is given by
\begin{equation}
     \mathcal{L}_V=\epsilon_Ve A'_\mu\bar \psi \gamma^\mu \psi~.
\end{equation}
where $\psi$ is the electron field.

Just like for vector bremsstrahlung \cite{Tsai:1986tx,Bjorken:2009mm}, when $m_a \gg m_e$ there is a kinematic preference for producing energetic, forward-peaked bosons, particularly when the electron energy is much larger than the axion mass. The resulting differential cross-section is difficult to sample efficiently. To address this, we use the \texttt{PETITE} package~\cite{Blinov:2024pza} which employs a histogram approximation to the differential cross-section, that is used for fast importance sampling of the exact one. The approximation is provided by the \texttt{VEGAS} integrator trained using the dark photon matrix element~\cite{Lepage:2020tgj}. We use \texttt{PETITE} to sample the $V$ phase space and then reweigh by the full ratio of matrix elements squared (see \cref{app:weights} for details); this yields Monte Carlo (MC) samples that are distributed according to the correct ALP's phase space. 

For electron beams, bremsstrahlung and photon fusion compete as the primary production mechanism. Technically, the amplitudes for the two mechanisms should be summed coherently since the initial and final states are the same. However, as we have shown above, the two processes have little phase space overlap: photon fusion is enhanced at small axion energies due to the soft divergence of EPA flux, while bremsstrahlung is peaked for high-energy forward ALPs. Furthermore, as we will see below, subsequent shower-induced reactions entirely dominate the sensitivity projections. Therefore, we sum the two processes independently when computing the primary production yields.

\paragraph{\bf Resonant annihilation:} Positrons produced in the electromagnetic shower can annihilate resonantly on atomic electrons, 
\begin{equation}
    e^+ e^-\rightarrow a (\gamma)~.
\end{equation}
Note that we have allowed for the emission of a photon to bring the incident electron onto the resonant peak, i.e., radiative return. 

In \texttt{PETITE}, the analogous resonant annihilation process $e^+e^- \to V(\gamma)$ involving a dark vector in the final state is implemented using a framework that combines a QED parton distribution function for initial state radiation with a model for atomic binding effects. The treatment of atomic binding begins with the tree-level cross-section for a positron of momentum $\bm{k}$ annihilating an electron in a fixed atomic orbital, which is given by
\begin{align}
\sigma^{(0)}(k) =& \frac{1}{4m_e\sqrt{\omega^2 - m_e^2}}\int \frac{\dd^3p}{(2\pi)^3} \frac{1}{2E_e} \frac{1}{2E_V}\\&\times (2\pi)\delta(\omega + m_e - \varepsilon_A - E_V)|\psi_A(p)|^2 \langle|\mathsf{M}_V|^2\rangle \nonumber
\end{align}
where $|\psi_A(p)|^2$ is the bound-state electron wave-function, $\varepsilon_A$ is the binding energy, $E_V$ is the energy of the final-state vector and $\langle|\mathsf{M}_V|^2\rangle$ is the matrix element calculated for free-electron states. For many-electron atoms, \texttt{PETITE} approximates the total momentum distribution by modeling it as a sum of $1s$ hydrogenic orbitals. This choice enables an analytic calculation of the cross-section, which is then convolved with the leading-order $e^- \to e^-\gamma$ splitting function to account for initial state radiation. This method provides a computationally efficient framework for MC sampling while capturing the essential physics of resonance broadening due to the momentum of bound atomic electrons. 
The dark vector sample is then simply converted into an axion by reweighing it by the free-electron amplitude for the production of a pseudoscalar of the same mass (see~\cref{app:weights} for details). 

The cross-section is highly peaked near the resonant energy 
\begin{equation}
    E_{\rm res} \simeq \frac{m_a^2}{2 m_e}~, 
\end{equation}
and can dominate the axion flux. This means that resonant annihilation typically leads to flux that is dominated by $E\approx E_{\rm res}$. Depending on the mass of the axion, this flux can have modest or very large boosts. We direct the reader to~\cite{Zhou:2024aeu,Plestid:2024xzh} and references therein for a detailed discussion of the formalism for resonant production (see also \cite{Marsicano:2018glj,Marsicano:2018glj,Nardi:2018cxi,Marsicano:2018krp,Celentano:2020vtu,Arias-Aragon:2025xcc}).

\paragraph{\bf Compton-like scattering:} High energy photons can scatter off atomic electrons to produce an axion and an electron in the final state, 
\begin{equation}
    \gamma e^- \rightarrow a\,e^- ~.
\end{equation}
When considering a free electron at rest, this process requires a minimum axion energy:
\begin{equation}
    E_{\rm min} \simeq \frac{m_a^2}{2m_e}~.
\end{equation}

In \texttt{PETITE}, dark Compton scattering ($\gamma e^- \to e^- V$) is treated analogously to the resonant annihilation channel. The production cross-section is computed by convolving the QED splitting function $P_{\gamma \to e}(x)$ with the cross-section for resonant annihilation of the intermediate positron on a bound atomic electron, utilizing the same atomic framework as the direct annihilation channel described above \cite{Zhou:2024aeu}. The integration over the momentum fraction, $x$, of the intermediate positron averages over the narrow resonance condition. This results in a broad energy spectrum for the final-state vector, in contrast to the sharply peaked spectrum from direct $e^+e^-$ annihilation, with 
\begin{equation} 
\label{eq:dark_compton_atomic}
\sigma(k) = \langle|\mathsf{M}_V|^2\rangle \times \frac{2}{3m_e\Lambda} \frac{1}{k^2} \times \frac{\beta}{4} \mathcal{J}(a, b),
\end{equation}
where $\bm{k}$ is the momentum of the incident photon, $\Lambda=Z_{\rm eff}\,\alpha \,m_e$, $\beta=2\alpha/\pi\left(\log(s/m_e^2)-1\right)$, and $\mathcal{J}(a, b)$ is defined in Appendix A of \cite{Zhou:2024aeu}.

The sampled dark vector events are then re-weighted to model axion production by applying the ratio of the squared matrix elements, as described in \cref{app:weights}.
As the Compton-like channel provides a subdominant contribution relative to resonant annihilation across the parameter space of interest, we sum their respective yields for brevity's sake.

Our list of production mechanisms is not exhaustive, but we believe that it captures all dominant production mechanisms over the parameter space we consider. For example, one can check that electromagnetically mediated meson decays such as $\pi^0 \rightarrow \gamma \gamma a$ or $\eta \rightarrow \gamma \gamma a$ will not compete with Primakoff production. Similarly, virtual Primakoff production (from a nearly-on-shell $\gamma^*$ emitted by a fast proton) is subdominant relative to on-shell Primakoff production \cite{Dobrich:2019dxc} (as we have explicitly verified).

\section{Monte Carlo Methods}
\label{sec:methods}

The expected number of detectable ALP events in a beam dump experiment depends on several sequential processes: ALP production in the target, survival during transport, decay within the observable decay region, geometric acceptance, and energy thresholds. Conceptually, if these processes were independent with uniform probabilities, the event rate could be written as:
\begin{equation}
    N_{\text{events}} \sim N_{\text{prim}} \times P_{\text{prod}} \times  P_{\text{decay}}\times P_{\text{geo}} \times P_{\text{Ecut}}~,
\end{equation}
where $N_{\text{prim}}$ is the number of particles-on-target (POT), $P_{\text{prod}}$ is the probability of producing an ALP, $P_{\text{decay}}$ is the decay probability within the decay pipe and/or detector volume, $P_{\text{geo}}$ is the geometric acceptance, and $P_{\text{Ecut}}$ represents energy threshold requirements.
However, this simplified picture breaks down in realistic simulations where each ALP candidate has different kinematics, production mechanisms, and acceptance criteria. Given the extremely small probabilities involved and the need to account for event-by-event variations, we instead calculate the event rate using a weighted MC approach. The simulation begins by generating a large sample of SM showers in the target using \texttt{PETITE}, initiated by the incident particle beam. From this shower, we generate a sample of ALP candidates, applying a series of weights corresponding to each physical process.

\paragraph{Production Probability}
This is the number of ALPs produced per primary particle. We treat every particle in the shower as a potential source and compute weighted events. For each shower particle, we generate ALP candidates using the appropriate differential cross-sections (see \cref{sec:production}). Each candidate $i$ is then assigned a production weight, $w_{\text{prod},i}$, equal to the total probability of its production:
\begin{equation}
    w_{\text{prod},i} \propto \frac{\sigma_a}{\sigma_{\rm SM}}w_{{\rm prim},i}\,,
\end{equation}
where $\sigma_a$ is the (total) cross-section for generating an axion, $\sigma_{\rm SM}$ is the cross-section of any other SM process, and $w_{{\rm prim},i}$ is the MC weight assigned to the parent/primary particle. For the specific details on how the weight is assigned for each of the production mechanisms described above, see Appendix~\ref{app:weights}. 

\paragraph{Decay Probability}
This is the probability that an ALP decays within the decay volume of length $L$, situated at a distance $d$ from the target. This is applied as a weight to each candidate $i$:
\begin{equation}
    w_{\text{decay},i} = \e^{-d/\lambda_i} \left(1 - \e^{-L/\lambda_i}\right),
    \label{eq:decay_weight}
\end{equation}
where $\lambda_i = (|\bm{p}_i|/m_a) c \tau$ is the lab-frame decay length of the candidate.

In our study, we first consider the di-photon decay mediated by the (effective) axion-photon coupling $\tilde c_{a\gamma\gamma}$, whose partial width is given by
\begin{equation}
\Gamma_{a\rightarrow\gamma\gamma}=\frac{\tilde{c}^2_{a\gamma\gamma}\alpha^2}{64 \pi^3} \frac{m_a^3}{f^2}\,,
\end{equation}
where $\tilde{c}_{a\gamma\gamma}$ is given in \cref{eq:cagg_rg_val}.
When electron couplings are present at tree-level, the partial width for $a\rightarrow e^+e^-$ can instead dominate,  
\begin{equation}
  \Gamma_{a\rightarrow e^+e^-} = \frac{c_{aee}^2}{2\pi} \frac{m^2_e m_a}{f^2} \sqrt{1-\frac{4 m_e^2}{m_a^2}}.
\end{equation}
We define the total lifetime $\tau$ in the particle's rest frame as:
\begin{equation}
    \tau\equiv\hbar\left(\Gamma_{a\rightarrow\gamma\gamma}+\Gamma_{a\rightarrow e^+e^-}\right)^{-1}.
\end{equation}
In the long-lived particle (LLP) limit ($\lambda_i \gg L,d$), the decay weight simplifies to $w_{\text{decay},i} \approx L/\lambda_i \propto 1/f^2$.

\paragraph{Geometric Acceptance}
This is the probability that the ALP's decay products intersect the detector. For the sake of simplicity, we consider only the ALP's trajectory. 
The geometry is characterized by the length of the decay volume, $L$, its distance from the target, $d$, and the cross-sectional area of the detector $A$. Due to relativistic kinematics, the ALP flux is typically collimated in the forward direction. Let $\theta_i$ be the angle with respect to the beam direction at which the $i-{\rm th}$ axion is produced. Let $\theta_{\rm acc}\approx \sqrt{A}/(2L+2d)$ be the angular acceptance of the detector. In our MC, the geometric weight is given by a binary filter applied to each candidate $i$:
\begin{equation}
    w_{\text{geo},i} = \begin{cases} 1 & \theta_i\le\theta_{\rm acc}~, \\ 0 & \text{otherwise}~. \end{cases}
\end{equation}
A more sophisticated treatment of the geometric acceptance of the decay products can be performed (see Appendix B of \cite{Zhou:2024aeu} for a detailed analysis of SHiP's geometric efficiency). However, this was shown to be relevant for low-energy downstream fluxes, which -- in our region of interest -- are mostly reduced by the energy acceptance threshold described below. 

\paragraph{Energy Acceptance} A decay is only considered a signal event if it can be successfully reconstructed. In practice, the experimental requirement is a minimum visible electromagnetic energy (e.g.\ from decay photons and electron-positron pairs) in the detector;
we implement this as a cut on the parent ALP energy $E_{\text{cut}}$. This acceptance criterion is applied as a step-function weight:
\begin{equation}
    w_{\text{Ecut},i} = \begin{cases} 1 & E_i\ge E_{\text{cut}} \\ 0 & \text{otherwise} \end{cases}\,,
\end{equation}
where $E_i$ is the energy of the ALP candidate. Following the discussion in \cite{Nico:chat,Zhou:2024aeu}, we set $E_{\text{cut}}=200$ MeV for SHiP (lower than the nominal $1~{\rm GeV}$). For BDX, we set $E_{\text{cut}}=300$ MeV in agreement with the observation made in \cite{Essig:2024dpa}.
We illustrate the effect of different energy cuts on the sensitivity curves in \cref{app:ecuts}.

\bigskip

In conclusion, the total number of events per POT is the sum of the final weights over all generated candidates:
\begin{equation}
\label{eq:MCNevts}
    \frac{N_{\text{events}}}{ N_{\text{prim}}} = \sum_{i} (w_{\text{prod},i} \times w_{\text{decay},i}\times w_{\text{geo},i} \times w_{\text{Ecut},i} ).
\end{equation}
We note that in the LLP limit, $w_{\text{decay}} \propto 1/f^2$. Since $w_{\text{prod}} \propto 1/f^2$, the total event rate in the LLP limit scales as $N_{\text{events}} \propto 1/f^4$. Therefore, in the LLP limit the \emph{shape} of the axion energy distribution can be rescaled by a factor $f^4$ to make it coupling-independent. In contrast, when the decay length is comparable to experimental length scales, the event weights become exponentially sensitive to $f$ (c.f. \cref{eq:decay_weight}).
\section{Experimental projections}
\label{sec:experiments}

\begin{figure*}[t]
\begin{ruledtabular}
\begin{tabular}{lcc}
\textbf{Parameter} & \textbf{SHiP} & \textbf{BDX}
\\
\hline
\noalign{\vskip 0.8ex}
Beam & 400 GeV protons & 10.6 GeV electrons \\
Target & Molybdenum & Aluminum \\
Integrated Exposure (particles-on-target) & $6\times 10^{20}$ POT & $1\times 10^{22}$ EOT \\
Distance to decay $d$ & $33.5~\mathrm{m}$ & $20~\mathrm{m}$ \\
Decay pipe/fiducial volume $L$ & $50~\mathrm{m}$ & $2.95~\mathrm{m}$ \\
Transverse acceptance area $A$ & $4~\mathrm{m}\times 6~\mathrm{m}$ & $0.50~\mathrm{m}\times 0.55~\mathrm{m}$  \\
Energy acceptance threshold $E_{\rm cut}$ & $200~\mathrm{MeV}$ & $300~\mathrm{MeV}$ \\
\end{tabular}
\end{ruledtabular}
\caption{Benchmark experimental parameters used in our SHiP and BDX projections. For BDX there is no decay pipe; detectable decays must occur
within the calorimeter fiducial volume.}
\label{fig:exp-benchmarks}
\end{figure*}

In what follows, we focus on two representative experiments, SHiP and BDX, though our methodology applies broadly to other beam dump experiments. We have developed the publicly available code \href{https://github.com/spatrone/ALPETITE}{\faGithub~\texttt{ALPETITE}}, which converts electromagnetic cascades and dark vector fluxes from \texttt{PETITE} simulations into ALP event predictions for arbitrary beam dump configurations. The modular design allows users to easily modify experimental parameters (geometry, POT, acceptance thresholds) and theoretical inputs (couplings, masses) without the need of regenerating Monte Carlo samples. 

The SHiP experiment~\cite{Aberle:2839677,Albanese:2878604} will use a 400 GeV proton beam incident on a molybdenum target with the goal of collecting $6\times 10^{20}$ POT. An axion is detectable if it decays visibly in some region of space after the beam dump. The version of
the experiment that was approved in March 2024 is more
compact than the original design. For long-lived particle signals considered here, the relevant detector subsystem is the hidden sector decay spectrometer (HSDS). The new decay pipe starts at $d = 33.5~{\rm m}$ from the target and is $L = 50~{\rm m}$ long, followed by a charged particle tracker, timing detector, electromagnetic and hadronic calorimeters. The tracker has an aperture of $4~{\rm m} \times 6~{\rm m}$, which determines the angular acceptance of potential decays.
Following \cite{Jerhot:2022chi}, we simulated the $N_\gamma$ primary on-shell photons arising from $\pi^0$, $\eta$, $\eta'$ di-photon decays by using \texttt{PYTHIA 8.2} \cite{Sjostrand:2014zea} with the \texttt{SoftQCD:all} flag and a pomeron flux parametrization \texttt{SigmaDiffractive:PomFlux(5)}, for $N_p=10^5$ $pp$ collisions at a center-of-mass energy of 27.4 GeV. We randomly selected a subsample of the above list with energy above a certain threshold and assigned an initial MC weight to each primary photon equal to 
\begin{equation}
\label{eq:mcweight_photons}
    w_{\rm prim}=N_{\gamma}(E_\gamma>50~{\rm MeV})/N_p\times(1-\e^{-1})\approx 5.34\,,
\end{equation}
where we added the exponential factor as a conservative choice to consider only the $\gamma$ flux coming from the first interaction length.  The weight approximately matches the number of expected photons per POT ($\sim 5$) for SHiP in \cite{Magill:2018tbb}, Table I. For transparency, we show in App.~\ref{app:ship-photon-spectra} (Fig.~\ref{fig:ship-photon-sample-vs-parent})
the energy and transverse-momentum spectra of the sampled primary-photon list used to seed the SHiP cascade,
together with a diagnostic comparison to the full parent photon population.
Finally, we generated SM showers for the primary photons using \texttt{PETITE} and produced axion samples of various masses, assigning MC weights as described in \cref{sec:methods}.

Based on the current proposal~\cite{BDX:2016akw}, the BDX experiment at Jefferson Lab will use a 10.6 GeV continuous-wave electron beam incident on the Hall-A aluminum beam dump with $10^{22}$ electrons on target (EOT). The detector will be housed in a new underground facility located $d\sim 20~{\rm m}$ downstream of the dump, behind $\sim 6~{\rm m}$ of shielding. Its active volume will be a plastic-scintillator electromagnetic calorimeter, with fiducial dimensions $\sim 50~{\rm cm} \times 55~{\rm cm} \times 295~{\rm cm}$, corresponding to a frontal acceptance area of about $\sim 0.275~{\rm m}^2$. Unlike SHiP, BDX will have no decay pipe, so detectable ALP decays must occur within the detector's fiducial volume. Consequently, we use $\theta^{\rm BDX}_{\rm acc}\equiv\sqrt{A}/2d\sim0.013$. The benchmark experimental parameters used in our projections are summarized in~\cref{fig:exp-benchmarks}.

\subsection{Downstream fluxes}
\begin{figure*}[t]
    \centering
    \begin{minipage}[t]{0.48\textwidth}
    \centering
    \includegraphics[width=\linewidth]{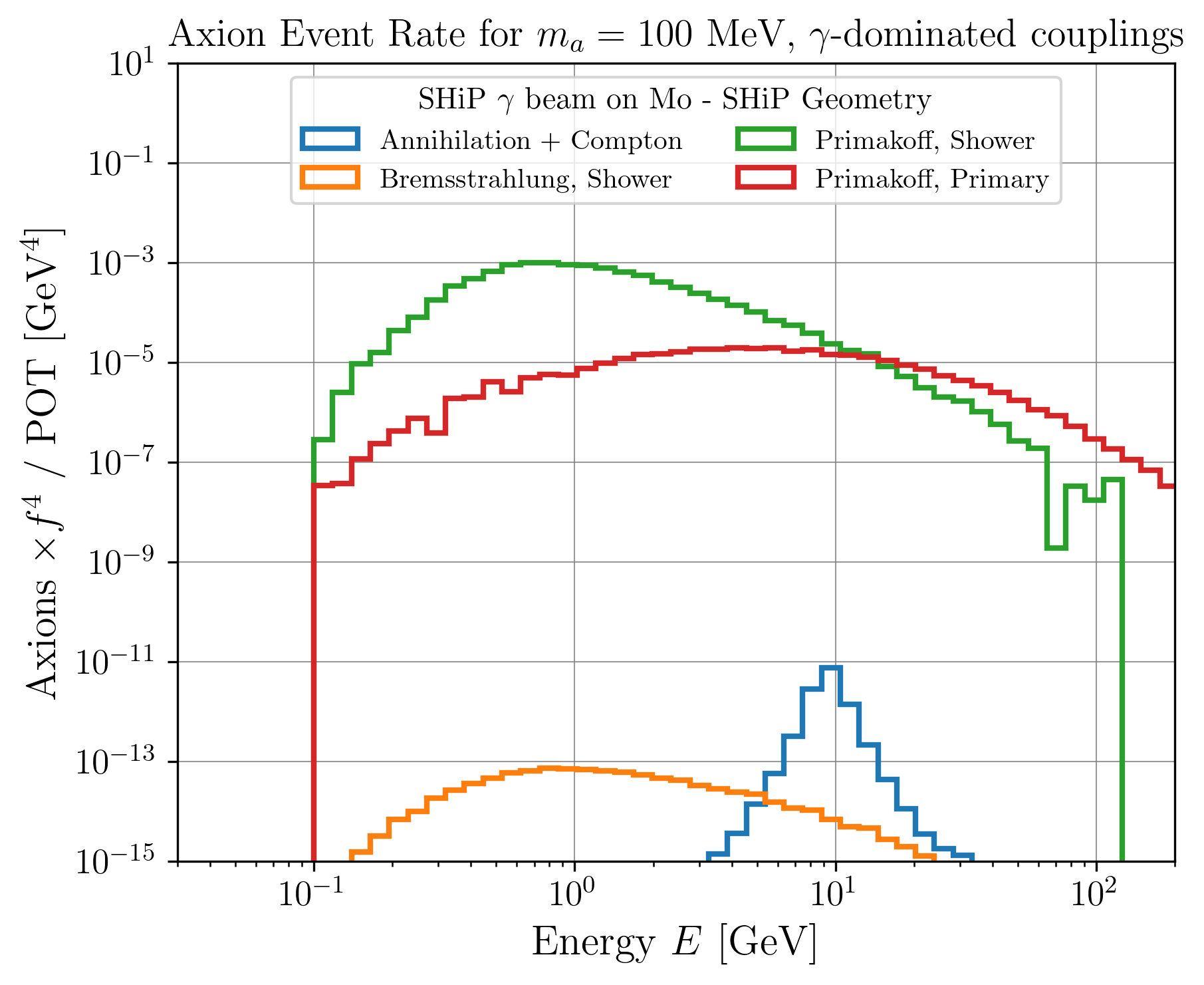}
    \caption{Event rate of axions per protons-on-target (POT), weighted by $f^4$, for an ALP with mass $m_a=100$ MeV and photon-dominated couplings. The setup simulates SHiP photons emerging from meson decays impinging on a molybdenum target, with the corresponding SHiP experimental geometry. The flux is calculated for ALPs directed towards the detector. The individual production channels are: Primakoff conversion from shower photons (green); bremsstrahlung from shower electrons and positrons (orange); $e^+e^-$ annihilation and Compton-like scattering of shower $e^\pm$ (blue); Primakoff conversion from the primary photon beam from neutral meson decays (red). \label{fig:ship_flux}}
    \end{minipage}
    \hfill 
    \begin{minipage}[t]{0.48\textwidth}
    \centering
    \includegraphics[width=\linewidth]{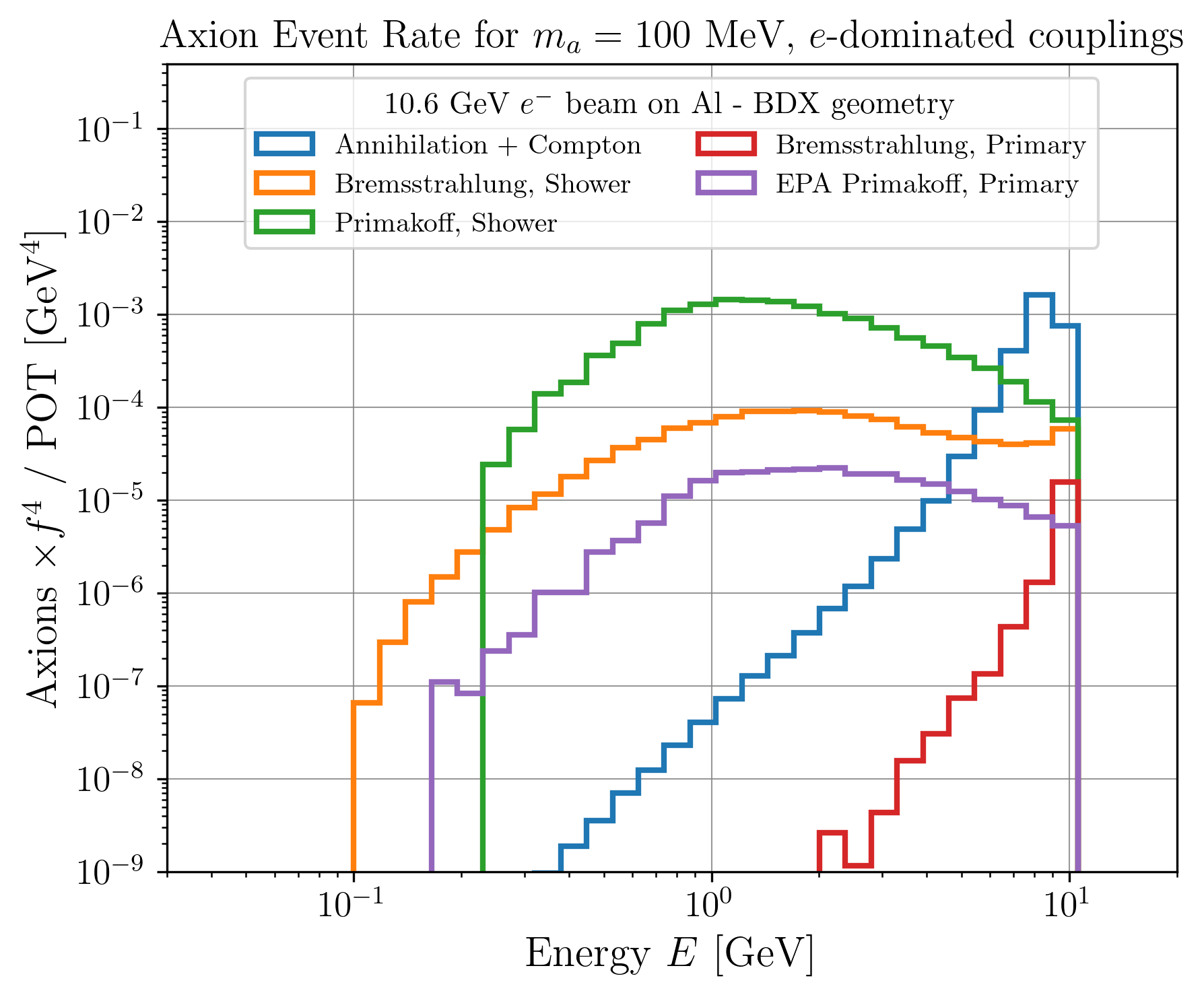}
    \caption{Event rate of axions produced per electrons-on-target (EOT), weighted by $f^4$, for an ALP with mass $m_a=100$ MeV and electron-dominated couplings. The setup simulates a 10.6 GeV electron beam on an aluminum target, with a geometry representative of the BDX experiment. The flux is restricted to ALPs with trajectories pointing towards the forward detector. The contributions from different production mechanisms are shown separately: Primakoff conversion from shower photons (green); bremsstrahlung from shower electrons and positron (orange); $e^+e^-$ annihilation and Compton-like scattering from shower particles (blue); bremsstrahlung (red) and photon fusion (purple) from the primary electron beam.
    \label{fig:bdx_flux}}
    \end{minipage}
\end{figure*}

\begin{figure*}[t]
    \centering
    \begin{minipage}[t]{0.48\textwidth}
    \centering
    \includegraphics[width=\linewidth]
    {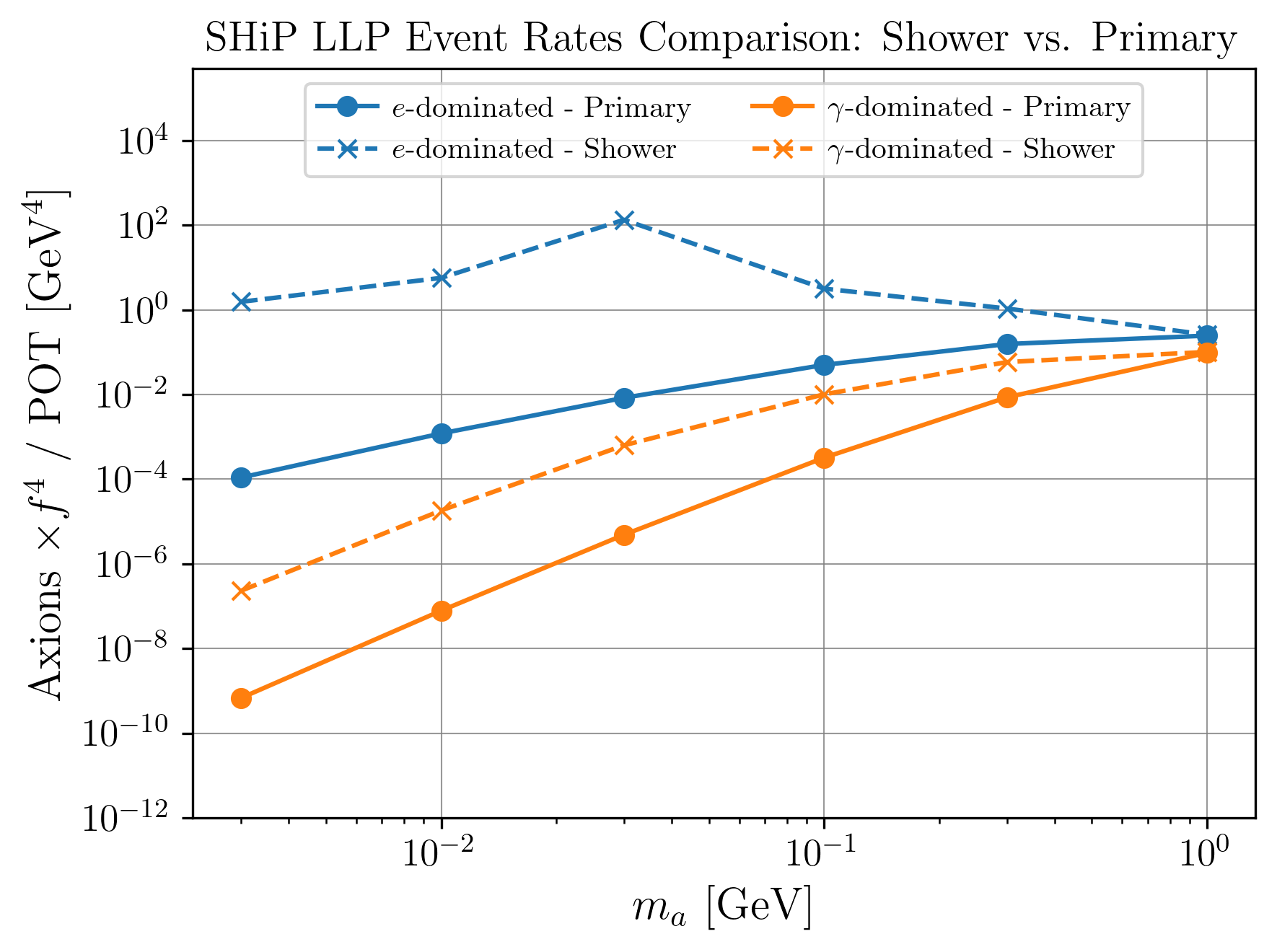}
    \label{fig:SHIP_flux_tot}
    \end{minipage}
    \hfill 
    \begin{minipage}[t]{0.48\textwidth}
    \centering
    \includegraphics[width=\linewidth]{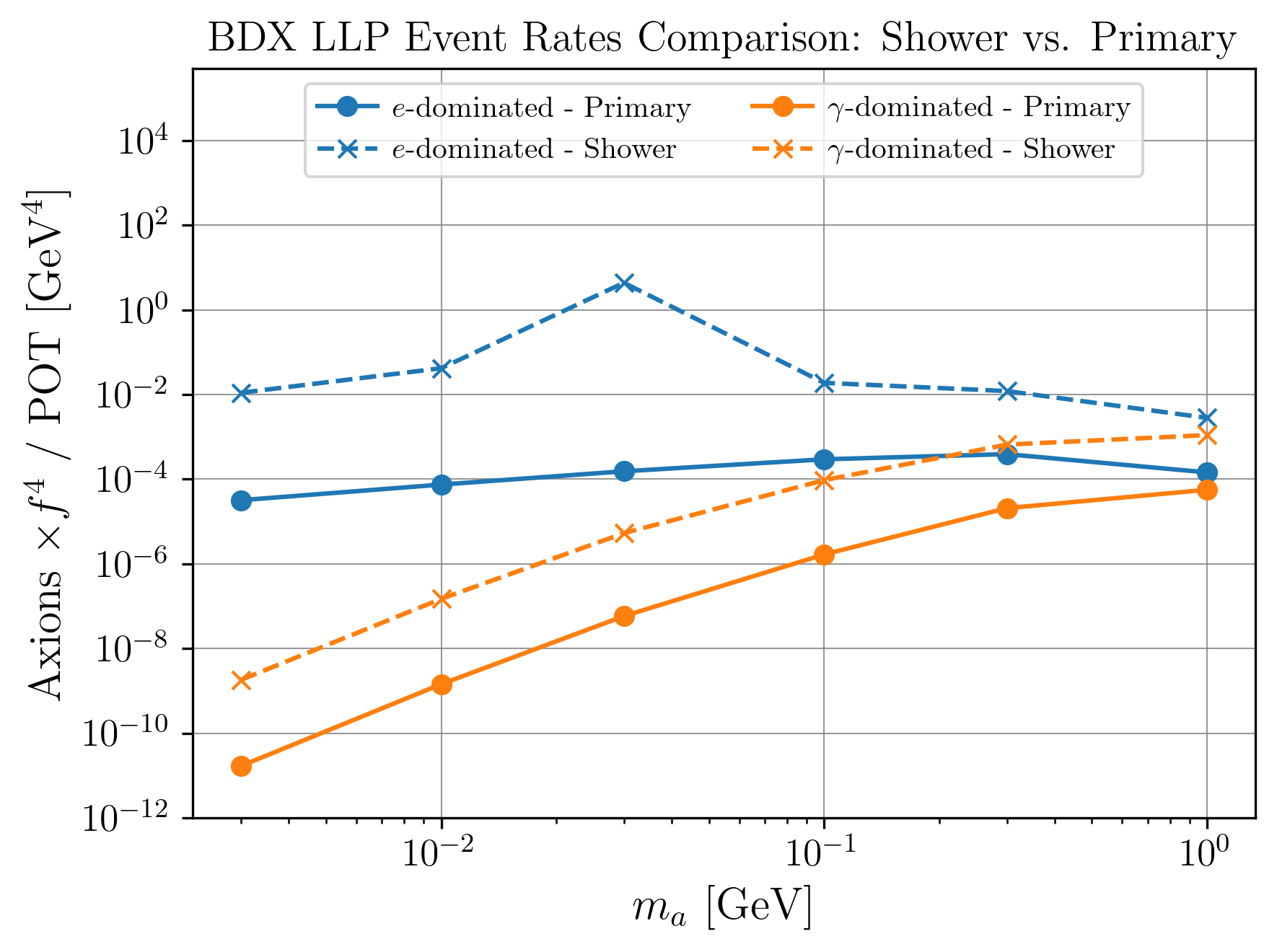}
    \label{fig:BDX_flux_tot}
    \end{minipage}
    \caption{Comparison between the primary-only and the shower-induced coupling-independent weighted event rates in the LLP limit for different axion masses for SHiP (\textbf{left}) and BDX (\textbf{right}). As above, the axion count has been weighted by decay (in the LLP limit), geometry, and energy acceptance. 
   In both panels, as the mass is increased, the ability of softer shower particles to produce axions is diminished, and production rates from the showers becomes smaller (in the case of SHiP) than the primary-only estimates for masses greater than $1~\text{GeV}$.}
    \label{fig:flux_comparison}
 \end{figure*}

We begin our study with an analysis of the predicted ALP coupling-independent event rates in the LLP limit for the two experimental setups described above. 

As an illustration, in \cref{fig:ship_flux,fig:bdx_flux} we present the individual contributions from each process described in \cref{sec:production} for $m_a=100~\text{MeV}$ ALP in the $\gamma$-dominated benchmark (SHiP) and the $e$-dominated benchmark (BDX). 
The spectra are histograms of the expected event rate as a function of energy. The quantity on the $y$-axis, ``Axions~$\times~f^4 / $ POT'', is the weighted event rate per POT for each energy bin, with the coupling dependence factored out. This is constructed by summing the total weights of all MC candidates that fall within that energy bin and setting $1/f=1$, such that the event rate depends only on the ALP mass and experimental setup.

Examining the plots, several salient features emerge.
First, Primakoff production dominates over electron bremsstrahlung across most of the parameter space in both the electron- and photon-dominated benchmarks. 
In the electron-dominated scenario, this occurs because the RG-induced photon coupling, \cref{eq:cagg_rg_val}, is of comparable size with respect to the electron one.
For $m_a = 100~\text{MeV}$, the peak Primakoff flux arising from the electromagnetic shower exceeds the primary-beam peak by $\mathcal{O}(10^2)$; consequently, we expect at least a factor of $3$ improvement in sensitivity to $1/f$ from this contribution alone in the LLP limit. Notably, even in the electron-dominated case, electron bremsstrahlung remains subdominant to Primakoff production once $m_a \gtrsim 100~\text{MeV}$. Resonant annihilation of positrons on atomic electrons yields competitive ALP fluxes at lower masses, with a spectrum peaked near $E_{\rm res} \simeq m_a^2/(2 m_e)$. 
There are two production mechanisms in an electron beam dump experiment (e.g.~BDX) from the primary $e^-$ beam: bremsstrahlung, mediated by the axion-electron coupling ($c_{aee}$), and photon fusion, mediated by the (effective) axion-photon coupling ($\tilde c_{a\gamma\gamma}$). The dominance of one process over the other is model-dependent. 
The ratio of their raw integrated fluxes, 
$\Phi$, without experimental cuts, is well approximated by the following expression, which we have confirmed via MC simulation:
\begin{equation}
\label{eq:flux_ratio}
\frac{\Phi_{\rm \gamma-fusion}}{\Phi_{\rm brem}}\sim \, \frac{\tilde c^2_{a\gamma\gamma}}{c^2_{aee}}\, \alpha\, \frac{m_a^2}{2m_e^2}\log^{-1}\left(\frac{m_a^2}{2m_e^2}\right)\,.
\end{equation}
For instance, in a scenario with equal couplings ($c_{aee}=~\tilde c_{a\gamma\gamma}$, e.g.~the $e$-dominated scenario of \cref{sec:alp_eft}), photon fusion surpasses bremsstrahlung to become the leading \emph{primary} process for axion masses $m_a\ge 20~\text{MeV}$.

In \cref{fig:flux_comparison}, we show the integrated ALP event rate in the LLP limit for both primary and shower-only particles, at several ALP masses. As in the histograms above, the yield is weighted by the decay probability, the detector energy threshold, and the angular acceptance. We render the result coupling-independent by factoring out the decay constant $f$ and normalizing to the particles on target. A flux enhancement of $\mathcal{O}(10^4)$ corresponds to an $\mathcal{O}(10)$ gain in sensitivity at small coupling. For both experiments and both benchmarks, the flux gain is pronounced -- especially at lower masses, where the broader phase space enables substantially more ALP production.

For SHiP, electromagnetic cascades enhance the flux by up to $\mathcal{O}(10^3)$ in the photon-dominated scenario and up to $\mathcal{O}(10^4)$ in the electron-dominated scenario at the lowest simulated mass ($3~\text{MeV}$). A comparably large enhancement also appears at higher masses ($30~\text{MeV}$) in the electron-dominated benchmark, primarily due to resonant annihilation.

For BDX, the flux enhancement is at least $\mathcal{O}(10^2)$ across all masses for both $e$-dominated and $\gamma$-dominated couplings, with the same resonant feature at $m_a = 30~\text{MeV}$ in the electron-dominated case. 

\subsection{Sensitivity projections}

 \begin{figure*}[t]
    \centering
    \begin{minipage}[t]{0.48\textwidth}
    \centering
    \includegraphics[width=\linewidth]
    {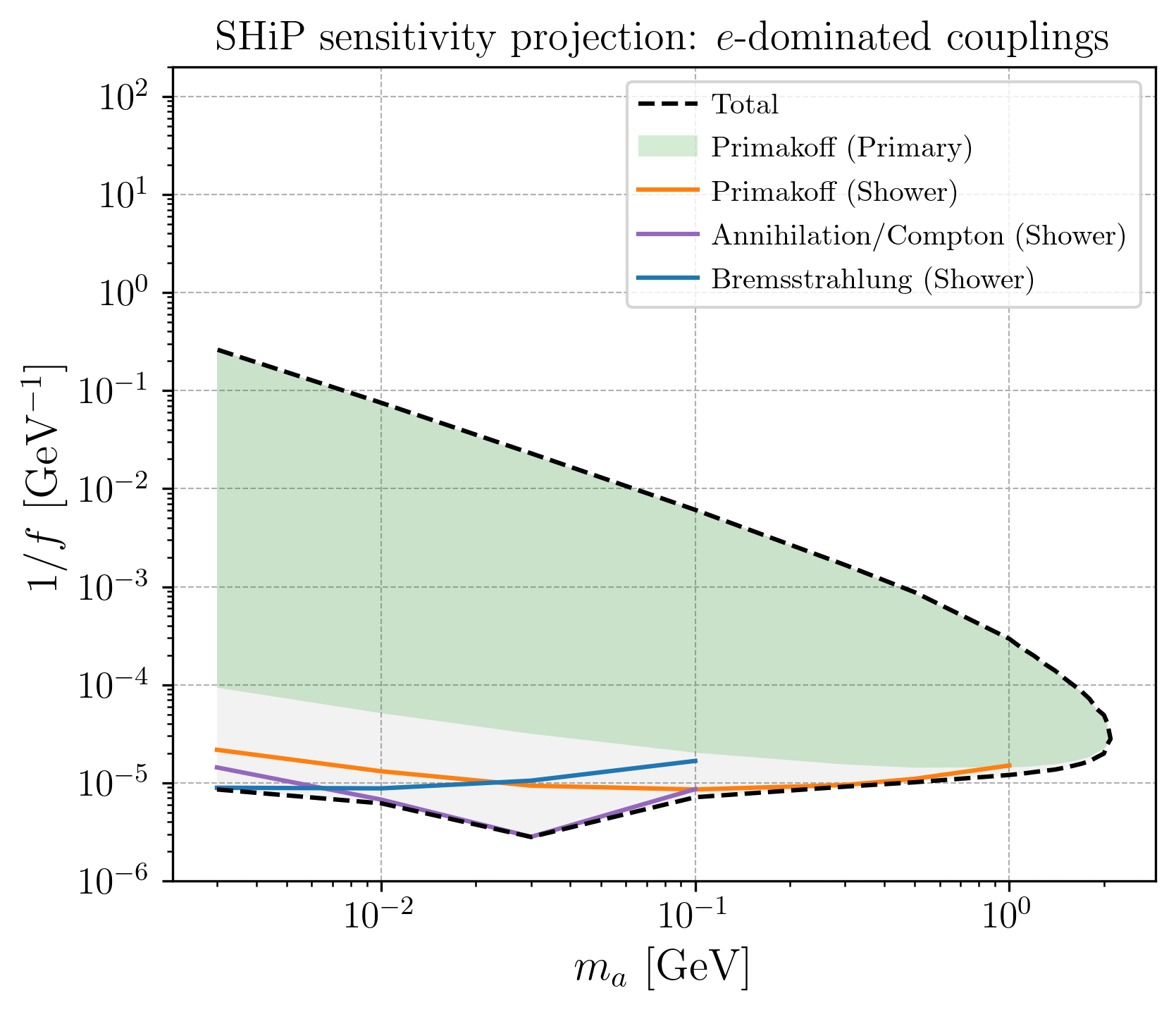}
    \label{fig:ship_sens_edom}
    \end{minipage}
    \hfill 
    \begin{minipage}[t]{0.48\textwidth}
    \centering
    \includegraphics[width=\linewidth]{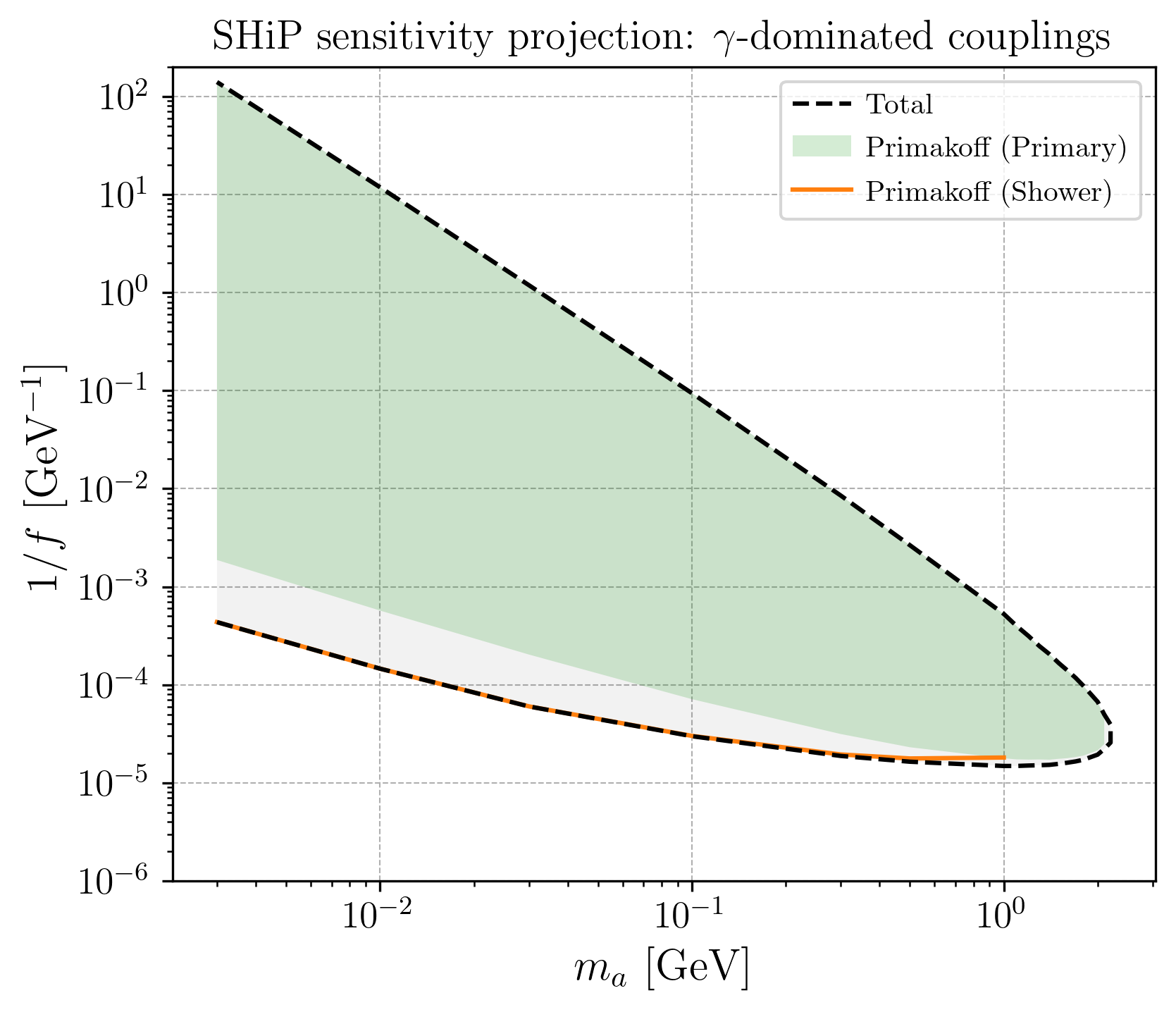}
    \label{fig:ship_sens_ydom}
    \end{minipage}
    \caption{Projected sensitivity for 5 signal events for the SHiP experiment for the electron-dominated (\textbf{left}) and photon-dominated (\textbf{right}) ALP coupling scenarios, assuming an integrated $6 \times 10^{20}$ POT and an energy acceptance threshold
$E_{\text{cut}}=200~\mathrm{MeV}$. The green shaded region shows the sensitivity for ALPs generated by primary photons arising from neutral meson decays. The solid lines show the individual contributions from shower-induced processes: Primakoff (orange), bremsstrahlung (blue), and resonant annihilation/Compton scattering (purple). The total sensitivity, combining primary and shower contributions, is shown by the dashed black line, with the gray shaded area indicating the full parameter space covered. For the photon-dominated case, we do not show the bremsstrahlung, and resonant annihilation/Compton scattering lines since they are sub-dominant with respect to all the other processes. 
    \label{fig:ship_sens} }
   \vspace{1cm}
    \begin{minipage}[t]{0.48\textwidth}
    \centering
    \includegraphics[width=\linewidth]
    {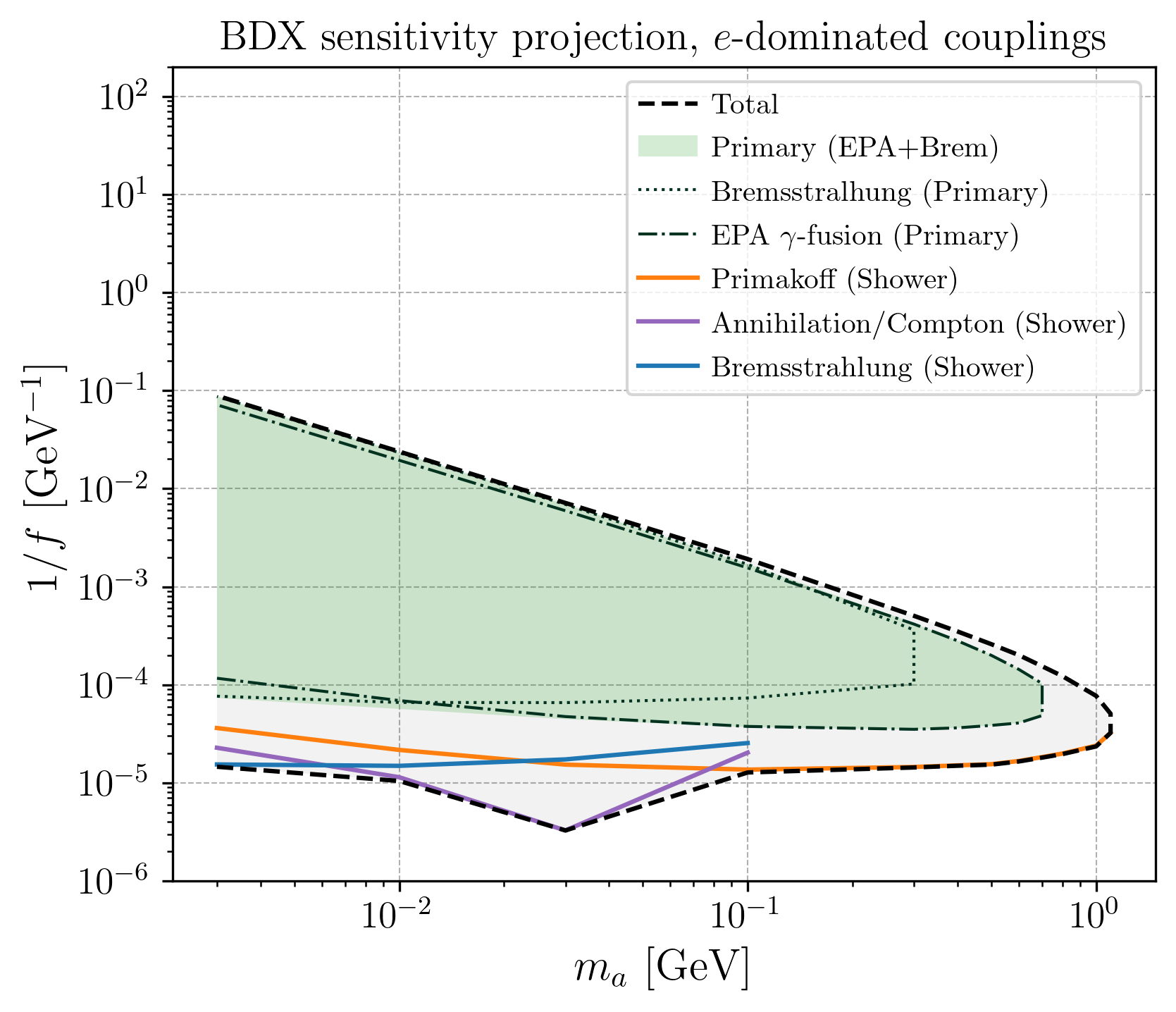}
    \label{fig:bdx_sens_edom}
    \end{minipage}
    \hfill 
    \begin{minipage}[t]{0.48\textwidth}
    \centering
    \includegraphics[width=\linewidth]{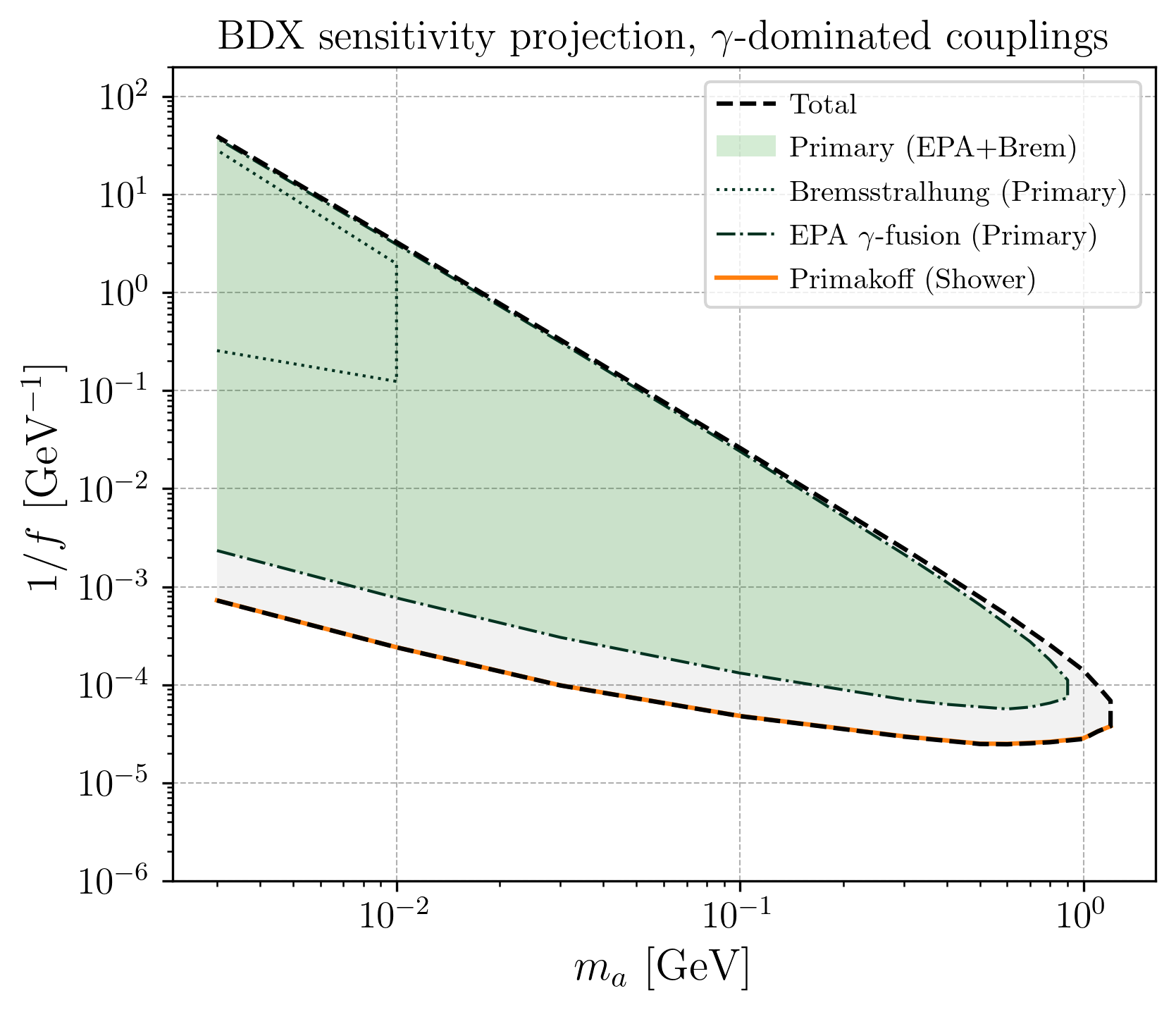}
    \label{fig:bdx_sens_ydom}
    \end{minipage}
    \caption{Projected sensitivity for 5 signal events for the BDX experiment for the electron-dominated (\textbf{left}) and photon-dominated (\textbf{right}) ALP coupling scenarios, assuming an integrated $10^{22}$ EOT and an energy acceptance threshold
$E_{\text{cut}}=300~\mathrm{MeV}$. The green shaded region represents the contribution from ALPs generated by the initial electron beam, and it is obtained by summing the contributions from the bremsstrahlung (dotted green line) and the EPA $\gamma$-fusion process (dot-dashed green line). The solid lines show the contributions from the subsequent electromagnetic shower: Primakoff (orange), bremsstrahlung (blue), and annihilation/Compton (purple). The total sensitivity is shown by the dashed black line, which encompasses the gray shaded region. For the photon-dominated scenario, we do not show the bremsstrahlung, and resonant annihilation/Compton scattering lines since they are sub-dominant with respect to the Primakoff channel.}
    \label{fig:bdx_sens}
 \end{figure*}

In \cref{fig:ship_sens,fig:bdx_sens}, we show how the inclusion of EM showers impacts the projected sensitivity of the experiments, accounting for a substantial increase in sensitivity in both electron-dominated and photon-dominated benchmarks.
The showers effectively multiply the number of secondary particles (photons, electrons, and positrons) within the target, opening up or enhancing various ALP production channels.
We estimate the sensitivity of each experiment using  \cref{eq:MCNevts}, and drawing $N_{\rm events}=5$ contours. The dependence on the assumed signal yield threshold (5, 10, and 100 events) is shown in App.~\ref{app:nsig}.
In \texttt{ALPETITE}, the function \texttt{compute\_and\_save\_sensitivities} can very easily compute sensitivity curves for any number of events desired and/or particles-on-target.

Many searches for Beyond the Standard Model (BSM) physics at SHiP are anticipated to be nearly background-free~\cite{Aberle:2839677}. The search for long-lived ALPs originating from the electromagnetic cascade is expected to share this characteristic. Indeed, preliminary analyses show no significant accumulation of background events at low track momenta~\cite{machado_santos_soares_2021_8shs1-8zw27}. Nevertheless, the precise background rate remains subject to final determination. 

A core design principle of BDX is the achievement of a nearly background-free environment. This is accomplished mainly by placing the detector far downstream of the dump, behind significant earth shielding, to suppress beam-related backgrounds. Consequently, searches for light dark matter are projected to be exceptionally clean. Extensive Monte Carlo studies have validated this approach, demonstrating that beam-related backgrounds and cosmogenic rays are reduced to negligible levels by the shielding and active veto systems, respectively~\cite{BDX:2016akw}. While these simulations are robust, the final background budget is subject to verification, and initial in-situ measurements from the BDX-MINI prototype have confirmed the effectiveness of the background mitigation strategies~\cite{Battaglieri:2020lds}.

For the SHiP experiment (\cref{fig:ship_sens}), which uses a proton beam, the primary source of high-energy SM particles capable of producing ALPs are the neutral mesons ($\pi^0, \,\eta, \,\eta'$) that decay into photons. 
In the photon-dominated scenario (right panel), the sensitivity is driven by the Primakoff process. As already pointed out in previous studies \cite{Dobrich:2019dxc, Jerhot:2022chi}, the primary photons from meson decays provide a strong baseline sensitivity (green region). However, as these photons initiate EM showers, they generate a much larger flux of lower-energy secondary photons, which in turn produce ALPs. This shower contribution (orange line) substantially improves the sensitivity, particularly for lower ALP masses ($m_a \lesssim 100$\,MeV), where the phase space for production by lower-energy photons is larger. 

The effect is even more dramatic in the electron-dominated scenario (left panel). Here, electron-coupling-dependent channels - like bremsstrahlung or resonant annihilation - are not available to primary photons. The sensitivity from primary photons alone (green region) cannot take advantage of the very efficient resonant annihilation channel (electron bremsstrahlung also plays an important role at lower masses). Once showers are included, the primary photons generate a cascade of secondary electrons and positrons. These secondaries then become a powerful source of ALPs via bremsstrahlung (blue line) and resonant annihilation/Compton (purple line). These shower-induced channels completely dominate the sensitivity for $m_a\le 100$~MeV, extending the reach in $1/f$ by more than an order of magnitude.

Because the shower particles have smaller energies, the resulting axion event yields are more sensitive to energy cuts compared to those coming from the highest energy primary particles. Nevertheless, in \cref{app:ecuts} we find that significant enhancements in sensitivity are obtained even when one requires $E_\text{cut} \sim \text{GeV}$.

For BDX (\cref{fig:bdx_sens}), the primary particles are electrons. In the electron-dominated scenario (left panel), the primary electrons can already produce ALPs via either bremsstrahlung or photon fusion (c.f. \cref{eq:flux_ratio} for an estimate of the relative importance of the two processes). The shower simulation enhances this by tracking the multiplication of electrons and positrons as they traverse the dump. Consequently, the shower-induced bremsstrahlung (blue line) and annihilation/Compton (purple line) contributions provide the dominant sensitivity for lighter masses. Notably, the Primakoff contribution from on-shell photons dominates for $m_a\ge100~\text{MeV}$, and already overcomes the (shower) bremsstrahlung contribution for $m_a\ge20~\text{MeV}$, in analogy to the primary case (c.f. \cref{eq:flux_ratio} and discussion below).

In the photon-dominated scenario for BDX (right panel of \cref{fig:bdx_sens}), the primary sensitivity curve is essentially dominated by the photon fusion process, across all the masses. Nevertheless, the electromagnetic cascade from the primary electrons generates a large number of photons, enhancing the ALP yield from the Primakoff process (orange line). As a result, the total sensitivity is increased by the shower-induced contribution, improving the reach across the entire mass range and at high masses.

\begin{figure*}[t]
   \centering
    \includegraphics[width=\linewidth]
    {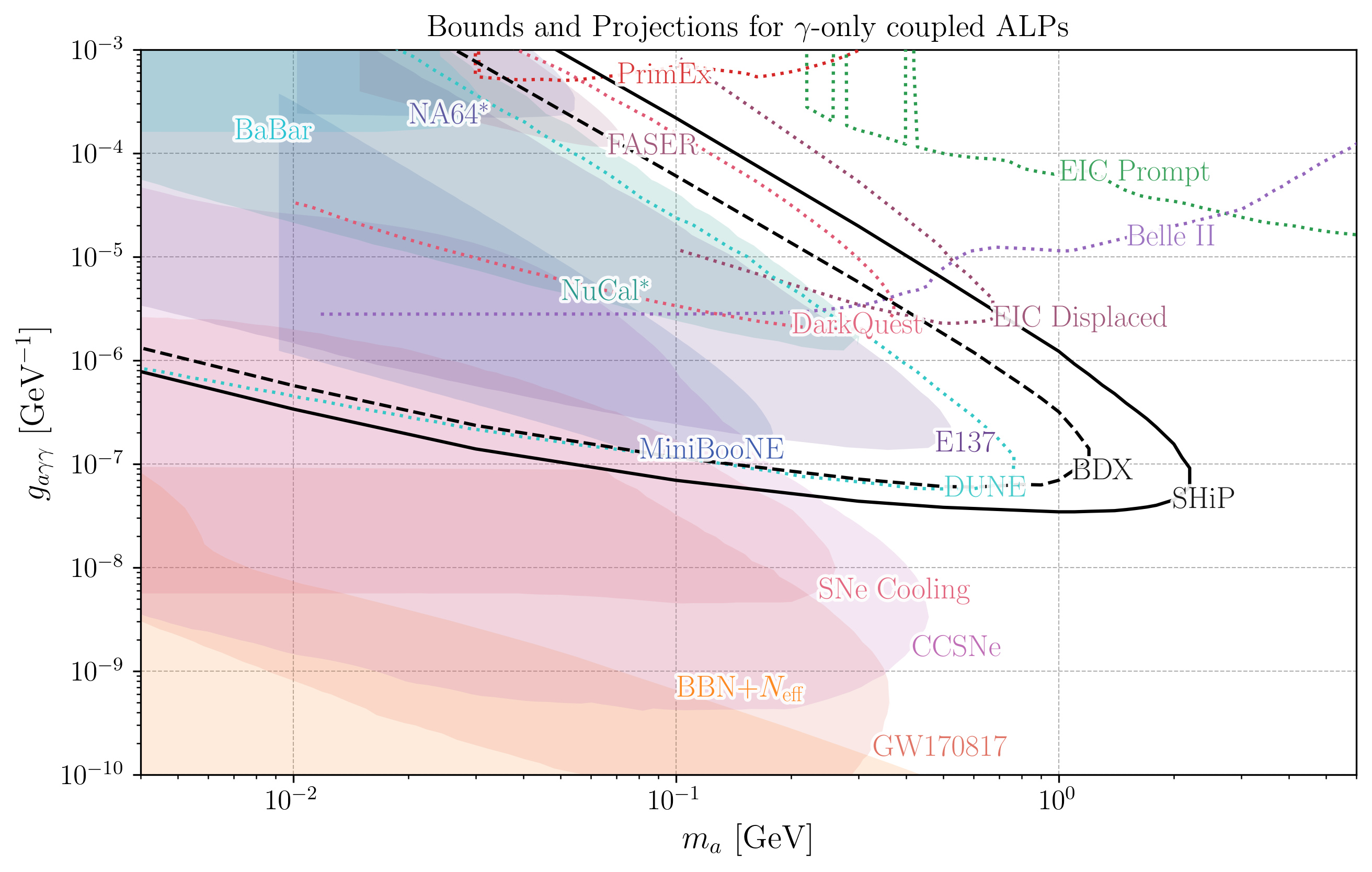}
    \caption{Comparison between our sensitivity projections for SHiP (black solid line) and BDX (black dashed line) with other relevant bounds (colored shaded regions) and projections (colored dotted lines) in the literature. We use $g_{a\gamma\gamma}\equiv f^{-1}c_{a\gamma\gamma}\alpha/\pi$, and we set $c_{aee}=0$ (the same conventions as \cref{fig:comparison-plot}). Starred bounds were computed without taking into account any enhancement from secondary production. Sources: PrimEx (projection only, current bound not shown) \cite{Aloni:2019ruo}, BaBar \cite{Dolan:2017osp}, NA64 \cite{NA64:2020qwq}, FASER (57.7 fb$^{-1}$) \cite{FASER:2024bbl}, EIC Prompt ($10~\text{fb}^{-1}$ line) and EIC Displaced  ($S_a=100,\, \left[10,100\right]\text{cm}$ line) \cite{Balkin:2023gya}, Belle II \cite{Acanfora:2024spi}, NuCal \cite{Blumlein:1990ay,Jerhot:2022chi, Blumlein:1991xh}, DarkQuest ($10^{18}$ POT) \cite{Blinov:2021say},  MiniBooNE Beam Dump \cite{MiniBooNEDM:2018cxm,Capozzi:2023ffu}, E137 \cite{Dolan:2017osp}, DUNE (GAr, 1 yr line) \cite{Brdar:2020dpr}, SNe Cooling and CCSNe ($1~\text{B}$ line) \cite{Fiorillo:2025yzf}, BBN$+N_\text{eff}$ \cite{Depta:2020wmr}, GW170817 fireball bound \cite{Diamond:2023cto}. LUXE-NPOD Phase-I \cite{Bai:2021gbm} projection curve is not shown since it is very close to DarkQuest.}
    \label{fig:bounds} 
\end{figure*}

In \cref{fig:bounds} we compare our projected sensitivities to other constraints in the literature. To ease comparison, we choose the photophilic scenario in which $c_{aee}=0$ and one considers only tree-level production involving the $c_{a\gamma\gamma}$ vertex. As we saw above, this scenario is essentially equivalent to the $\gamma$-dominated one in~\cref{sec:alp_eft}; it naturally arises when the ALPs are coupled only to photons in the UV (e.g.,~at the weak scale) and the induced electron coupling at lower scales is so small that it can be neglected. For consistency with the results quoted in the literature, we recast all the bounds in terms of the dimensionful coupling $g_{a\gamma \gamma} \equiv  f^{-1}(c_{a\gamma\gamma}\alpha/\pi)$, where we use the on-shell coupling $\alpha^{-1}=137.036$,  and take $c_{a\gamma\gamma}=1$ in converting between $f$ and $g_{a\gamma\gamma}$ (other choices of $c_{a\gamma\gamma}$ are obtained by trivial re-scalings). 

As emphasized in \cref{fig:comparison-plot}, the sensitivity derived herein for SHiP is stronger, and therefore covers more parameter space than past literature \cite{Beacham:2019nyx,Jerhot:2022chi}. The sensitivity curve for BDX derived here is, to the best of our knowledge, new. We find that the BDX collaboration will be able to probe currently untouched parameter space, reaching up to $m_a \gtrsim 1~{\rm GeV}$. 

At lower masses (below roughly $m_a \lesssim 200~ {\rm MeV}$) the regime in which we have derived enhanced sensitivity is largely covered by astrophysical constraints (in particular those stemming from supernovae). Nevertheless, the projections derived herein are complementary, and perhaps most importantly, can be obtained in a laboratory setting. Moreover, for $m_a\gtrsim 70~{\rm MeV}$ all of SHiP, BDX, and DUNE \cite{Brdar:2020dpr} are capable of probing axion-photon couplings $g_{a\gamma\gamma} \lesssim 2\times 10^{-7}~{\rm GeV}^{-1}$ corresponding to axion decay constants above the weak scale $f\sim 10~{\rm TeV}$.

Among the existing limits in the literature, some have incorporated electromagnetic cascades and others have not. For example, the E137 collaboration performed a Monte Carlo simulation of the electromagnetic shower in their aluminum and steel target station \cite{Bjorken:1988as}. To the best of our understanding, their simulation did not include the incoherent Primakoff-like production discussed above; however, the quoted limit ends at a low enough mass such that coherent production dominates over incoherent production and we expect their constraint to be reliable.\!\footnote{Only the long-lived limit was analyzed in \cite{Bjorken:1988as}, whereas the effect of attenuation due to decays was added later in \cite{Dolan:2017osp}.}  Re-analyses of CHARM \cite{CHARM:1985anb,Jerhot:2022chi} and NuCal \cite{Blumlein:1990ay,Jerhot:2022chi} have not included showers in the modelling of axion production in the target, and so we expect those experiments to have somewhat stronger reach than that shown in \cref{fig:bounds}. The effects of the electromagnetic cascade are less dramatic when recasting old analyses due to their higher energy cuts. For example, at E137 the observed signal in the downstream detector was required to have energy greater than $1~{\rm GeV}$ \cite{Bjorken:1988as}. This effectively truncates the shower despite a growing photon multiplicity at lower energies (see Fig.~14 of \cite{Bjorken:1988as}). 

In summary, we expect modest improvements to the sensitivity derived from old experimental data through the inclusion of electromagnetic showers, but the gains are somewhat limited by the typically large energy cuts applied in these old analyses. This lesson is important for upcoming experiments. As discussed in \cite{Zhou:2024aeu}, the SHiP experiment may be able to substantially extend its sensitivity by lowering its analysis thresholds. If BDX can achieve sub-300~MeV thresholds, this would also benefit their search for axion-like particles. 

\FloatBarrier

\section{Conclusions}
\label{sec:conclusion}
Axion-like particles are a well-motivated lamppost for new physics at very high scales. Being pseudo-Nambu-Goldstone bosons, they are naturally expected to be light, and when their decay constant is set by the high scale of new physics, they are further expected to be weakly coupled. When ALPs couple to hadrons (as is necessary for solutions of the strong-CP problem), they are efficiently searched for using direct production from meson decays or hadronic collisions. If, however, ALPs couple predominantly to leptons and/or electroweak gauge bosons in the ultraviolet, then leptonic and electromagnetic probes are best suited for their discovery. 

In this work, we have investigated both proton and electron beam dump experiments in detail. We find that the electromagnetic cascade will essentially always dominate the production of ``hadrophobic'' ALPs. Furthermore, enhancements relative to naive estimates based on the first radiation length are so large that, in fact, electromagnetic cascades can compete with hadronic production even if gluon couplings are only somewhat smaller than photonic couplings (say by a factor of five). Such a hierarchy can easily arise from a modest number of color-neutral fermions in the ultraviolet carrying electroweak quantum numbers.

Our main conclusions are as follows: when electron couplings are absent at tree-level, all electromagnetic phenomenology is dominated by the photon coupling. This statement holds essentially independently of the axion mass. By way of contrast, when electron couplings are present at tree-level they induce photon couplings which are capable of competing with electron-initiated processes. This crossover occurs around $m_a\simeq 100~{\rm MeV}$ with Primakoff production dominating above this mass and resonant annihilation dominating below it. 

In summary, our results demonstrate that electromagnetic cascades are not merely a correction, but an essential component for accurately estimating the sensitivity of beam dump experiments to ALPs, in both electron- and photon-dominated scenarios. This was well understood by the E137 collaboration \cite{Bjorken:1988as}, but it is often missing from sensitivity projections for upcoming experiments. This mirrors similar conclusions in the context of dark photon searches where resonant annihilation and the electromagnetic cascade more generally can dramatically enhance sensitivity at the weakest couplings accessible to the experiments \cite{Marsicano:2018glj,Marsicano:2018glj,Nardi:2018cxi,Marsicano:2018krp,Celentano:2020vtu,Blinov:2024pza,Zhou:2024aeu}.

Methodologically, our simulation framework goes beyond the simple one-dimensional approximation to shower development often used in the literature. By tracking the full electromagnetic cascade through \texttt{PETITE} - including transverse dynamics, multiple scattering, and realistic secondary production - we accurately predict angular distributions and thereby determine detector angular acceptances in realistic geometries.

While we have focused on SHiP and BDX, this lesson also applies to recasts of old experiments. For example, the sensitivity of NuCal \cite{Blumlein:1990ay} and CHARM \cite{CHARM:1985anb} has likely been underestimated in the literature \cite{Dobrich:2015jyk,Dolan:2017osp,Jerhot:2022chi}.  It would therefore be interesting to revisit the limits from these past experiments, including the missing production mechanisms discussed above. 

\section*{Acknowledgements}

We thank Leonardo Badurina and Yotam Soreq for useful conversations, and Kevin Kelly for help with \texttt{PETITE} and for feedback on the draft. We thank Edoardo Vitigliano for useful references.
RP and NB thank Kevin Kelly, Patrick J.~Fox, Pedro Machado, and Tao Zhou for collaboration on related work. SP thanks Federico Cima for useful discussions and Francesco Pio De Filippis for some graphic suggestions. RP was supported during this project by the Neutrino Theory Network under Award Number DEAC02-07CH11359. SP and RP were supported by the U.S. Department of Energy, Office of Science, Office of High Energy Physics under Award Number DE-SC0011632, and by the Walter Burke Institute for Theoretical Physics. NB acknowledges support from the Natural Sciences and Engineering Research Council of Canada (NSERC).

\newpage
\appendix
\section{Inelastic Primakoff-like scattering}
\label{app:inelastic_gain}
\begin{figure}[h!]
\vspace{-0.6cm}
    \centering \includegraphics[width=\linewidth]
    {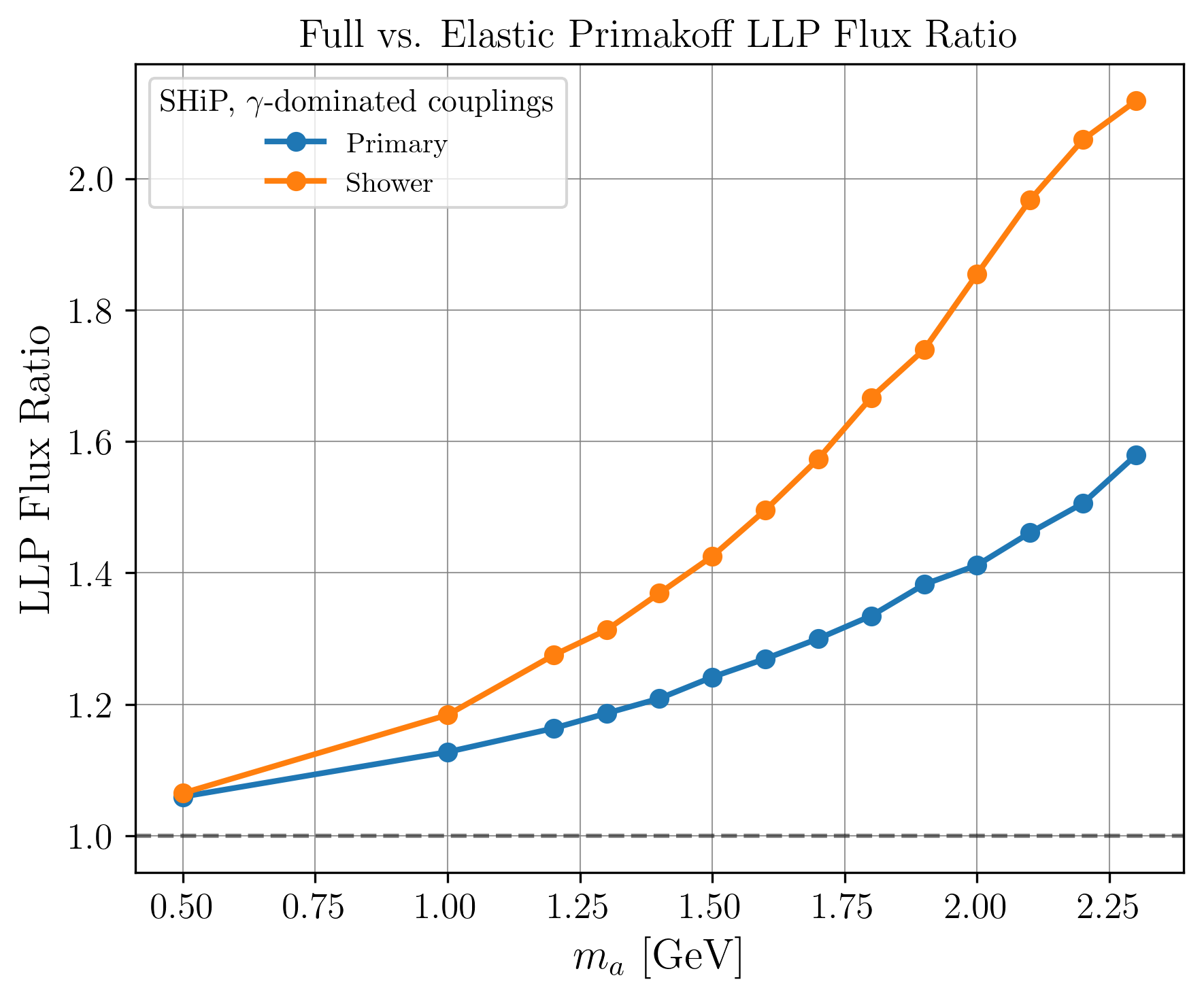}
\caption{Ratio of the projected downstream ALP flux (in the LLP limit) at SHiP calculated with and without the inelastic Primakoff form factor ($G_{2,\rm inel}$), for $\gamma$-dominated couplings. The enhancement is shown separately for ALPs produced by primary photons from neutral meson decays (blue) and secondary photons from electromagnetic showers (orange). The inclusion of inelastic scattering provides an $\mathcal{O}(1)$ enhancement to the flux, which becomes increasingly significant at higher ALP masses.}
     \label{fig:ship_inelastic}
\end{figure}

\bigskip

In this appendix, we assess the impact of including inelastic Primakoff-like scattering -- encoded by the inelastic nuclear form factor $G_{2,\rm inel}$ from \cref{eq:G_inel} -- on the predicted ALP flux at SHiP. This production channel is often neglected in the literature but can provide a non-negligible contribution to the total rate. Figure~\ref{fig:ship_inelastic} quantifies this effect by showing the ratio of the projected ALP flux calculated with the full form factor to that calculated with the coherent term alone ($G_{2,\rm inel}=0$). 

For both primary photons from meson decays and secondary photons from electromagnetic cascades, the inelastic channel provides an $\mathcal{O}(1)$ enhancement. This effect is most pronounced at higher ALP masses, as the larger momentum transfer required for production preferentially probes the inelastic regime where the nucleus breaks up. Because this enhancement depends fundamentally on the effective ALP–photon coupling, a comparable gain is expected in the $e$-dominated scenario. While a factor-of-a-few increase in flux leads to only a modest shift in the final sensitivity contours, it is a crucial correction for accurately predicting event rates, particularly for heavier ALPs and for experiments using lighter target nuclei.


\section{Monte Carlo production weights}
\label{app:weights}
In the following, we describe how we computed the production weights $ w_\text{prod}$ for each process described in Sec.~\ref{sec:production}. For initial particles with primary weights $w_{\text{prim}}$ (e.g., SHiP primary photons), the production weight is simply multiplied by the value of the primary weight given. 

\bigskip

\paragraph{Primakoff}
Primakoff production is controlled by nuclear and atomic (squared) form factors $G_{2,\rm el}(\absQ^2), G_{2,\rm inel}(\absQ^2) \leq1$, defined above in \cref{eq:G_el,eq:G_inel}. It is therefore convenient to compute the relevant cross-section where the sum of these two factors is set to unity and subsequently apply a ``screening weight'' i.e., 
\begin{equation}
    w_{\text{prod},i} = \frac{\sigma_P}{\sigma_{\rm SM}} \left[G_{2,\rm el}(\absQ^2) + \frac{G_{2,\rm inel}(\absQ^2)}{Z}\right]~,
\end{equation}
where $\sigma_P$ is computed for an infinitely heavy, point-like, and unscreened nucleus of charge $Z$. We will refer to this as the ``Born-Primakoff cross-section,'' in formulae
\begin{align}
    \sigma_P=&-\qty(\frac{1}{f})^2\frac{Z^2 \alpha^3 |\tilde{c}_{a\gamma\gamma}|^2}{32 \pi^2\omega^2}\int^{Q^2_{+}}_{Q^2_{-}} {\dd Q^2} \,\times \nonumber\\ &\left(Q^4+2Q^2(m_a^2-2\omega^2)+m_a^4\right)/Q^4
    \label{eq:sigma_born_integrated}
    \\\equiv& \frac{\alpha g^2_{a\gamma\gamma} Z^2}{8}
    \bigg[
        \left(1+\beta^2\right) \log\left(\frac{1+\beta}{1-\beta} \right)- 2\beta
    \bigg]~,\nonumber
\end{align}
where $g_{a\gamma\gamma}\equiv \alpha \tilde{c}_{a\gamma\gamma}/\pi f$, $\beta\equiv|\bm{p}_a|/\omega=\sqrt{1-m_a^2/\omega^2}$ and the kinematic boundaries are $Q^2_\pm=(\omega\pm|\bm{p}_a|)^2$.
The $Q^2$ of the axion is sampled according the Born-Primakoff differential cross-section (given in~\cref{eq:dsigma_primakoff}). This ensures that the expected value for $w_{\text{prod}}$ corresponds to the full screened cross section. In formulae:
\begin{equation}
     \langle w_{\text{prod},i} \rangle = \frac{\sigma_P}{\sigma_{\text{SM}}} \dfrac{\int \dd Q^2\dfrac{\dd \sigma_P}{\dd Q^2 }F(Q^2) } {\int \dd Q^2 \dfrac{\dd \sigma_P}{\dd Q^2 } }\equiv \frac{\sigma^{\text{screened}}_P}{\sigma_{\text{SM}}}~,
\end{equation}
where $F(Q^2)\equiv G_{2,\rm el}(\absQ^2) + G_{2,\rm inel}(\absQ^2)/Z$.

For an incident electron beam, the photon fusion process $e^-Z\rightarrow e^-Za$ is simulated using the Equivalent Photon Approximation (EPA). From each primary electron of energy $E_e$, a fixed number of $n_{\rm samples}$ virtual photons are drawn to represent the EPA flux. The energy fraction of each photon, $x\equiv\omega/E_e$, is sampled from the probability distribution proportional to the EPA spectrum, $\dd N_\gamma / \dd x$, defined in~\cref{eq:epa_spectrum_x}. This sampled photon is then fed to the Primakoff production pipeline described above. The primary weight assigned to each photon is the total integrated photon flux per electron, divided by the number of samples: 
\begin{equation}
    w_{\rm EPA} = \frac{1}{n_{\rm samples}} \int^{x_{\rm max}}_{x_{\rm min}} \dd x \, (\dd N_\gamma / \dd x)\approx \frac{0.42}{n_{\rm samples}},
\end{equation}
where we used $E_e=10.6~\text{GeV}$, $x_{\rm min}=50~\text{MeV}/E_e$, and $x_{\rm max}=0.99$ for BDX simulation.
In addition, the virtuality of the photon, $Q^2$, is sampled log-uniformly between its kinematic limits, $Q^2_{\rm min}\equiv m^2_ex^2/(1-x)$ and $Q^2_{\rm max}\equiv x^2E_e^2$. While the on-shell approximation is used for the Primakoff cross-section calculation, the sampled virtuality is kinematically relevant as it dictates the scattering angle of the equivalent photon and is crucial for applying angular acceptance criteria. The final MC weight for the axion produced by photon fusion is the product of the two, $w_{\rm EPA} \times w_{\text{prod},i}$.

\bigskip

\paragraph{Dark Bremsstrahlung}
The kinematics of axion production by bremsstrahlung is identical to dark vector production; moreover, the squared matrix elements are similar (c.f., \cref{eq:R-matrix-elements}). This means that we can simply rescale the weight $w_V$ of dark vectors sampled by \texttt{PETITE} by the ratio of differential cross-sections:
\vspace{-0.1cm}
\begin{equation}
     w_{\text{prod},i} = \frac{\dd^3\sigma_a}{\dd^3\sigma_V}\, w_V~.
\end{equation}
The differential cross-section used in \texttt{PETITE} is given by Eq.~(10) in \cite{Gninenko:2017yus} (which makes use of results from \cite{Liu:2016mqv,Liu:2017htz}),
\vspace{-0.2cm}
\begin{equation}
    \begin{split}
\frac{\dd^3\sigma}{\dd x \, \dd \cos\theta \, \dd t} &= \frac{\epsilon_V^2 \alpha^3 |\bm{k}| E_0}{\left|\bm{p}\right| \left|\bm{k}-\bm{p}\right|} ~\frac{G^{\rm tot}_2(t)}{t^2} \\
& \hspace{0.1\linewidth} \times \left( \frac{1}{2\pi} \int_0^{2\pi} \dd\phi_q \frac{\left| A_{V} \right|^2}{8M^2} \right)~,
    \end{split}
\end{equation}
and it is derived by the following interaction Lagrangian
\begin{equation}
\label{eq:VeeLinteraction}
    \mathcal{L}_V=\epsilon_Ve A'_\mu\bar \psi \gamma^\mu \psi~.
\end{equation}
In more detail, $\frac{1}{2\pi}\int_0^{2\pi} \dd\phi_q \left| A_{V} \right|^2$ is given by 
\begin{equation}
    A_{V}^{(0)} + Y \cdot A_{V}^{(1)} + \frac{1}{W^{1/2}} A_{V}^{(-1)} + \frac{Y}{W^{3/2}} A_{V}^{(-2)}~,
\end{equation}
where the different contributions are defined in Eqs. (15)-(18) of \cite{Gninenko:2017yus} and the equations above.
The differential cross-section for the axion is given by making the substitution $A^{(i)}_V \rightarrow A^{(i)}_a$ in the $\phi_q$ integral, as follows
\begin{widetext}
\begin{flalign}
    &A^{(1)}_V \rightarrow\; A^{(1)}_a = \frac{4M^2}{\tilde{u}}~, &\\
    &A^{(0)}_V \rightarrow\; A^{(0)}_a = \frac{4M}{\tilde{u}^2} \left[ -4 E_0^2 M m_a^2 (x - 1)^2 + 2 E_0 t (m_a^2 - x(m_a^2 + \tilde{u})) + M (m_a^2 t + 2 \tilde{u}^2) \right]~,& \\
    &A^{(-1)}_V \rightarrow\; A^{(-1)}_a = \frac{4M}{\tilde{u}} \left[ E_0^2 (8 M m_a^2 (x - 1) - 4 M t x^2) - 2 E_0 t (m_a^2 (x - 2) + \tilde{u} x) + M (2 m_a^2 t + \tilde{u}^2) \right]~, &\\
    &A^{(-2)}_V \rightarrow\; A^{(-2)}_a = 4M \left[ m_a^2 (-4 E_0^2 M + 2 E_0 t + M t) \right]~,&
\end{flalign}
\end{widetext}
where the effective interaction lagrangian we are using is 
\begin{equation}
\label{eq:aeeLinteraction}
    \mathcal{L}_a=\epsilon_ae a \bar\psi \gamma_5 \psi, 
\end{equation}
with $\epsilon_a\equiv 2m_e c_{aee}/ef$, being $c_{aee}/f$ the coupling in \cref{eq:lagrangian}. Recall that the derivative (\cref{eq:lagrangian}) and the pseudoscalar (\cref{eq:aeeLinteraction}) interactions are equivalent at leading order in $a/f$~\cite{ParticleDataGroup:2024cfk}.

\bigskip

\paragraph{Annihilation and Compton}
In these cases, since both processes have cross-sections proportional to the spin-averaged square amplitude $\langle|\mathsf{M}_V|^2\rangle$, the Monte Carlo weights of the dark vectors $w_V$ are just rescaled by the axion cross-sections
$\langle|\mathsf{M}_a|^2\rangle$, in formulae
\begin{equation}
    w_{\text{prod},i}=\frac{\langle|\mathsf{M}_a|^2\rangle}{\langle|\mathsf{M}_V|^2\rangle}w_V=\frac{\epsilon_a^2}{\epsilon_V^2}\frac{m^2}{2(m^2 + 2m_e^2)}w_V,
\end{equation}
where $m=m_V=m_a$, and we used the same interaction lagrangians in Eqs.~\eqref{eq:VeeLinteraction} and \eqref{eq:aeeLinteraction}.

\begin{figure*}[t]
    \begin{minipage}[t]{0.48\textwidth}
    \centering
    \includegraphics[width=\linewidth]
    {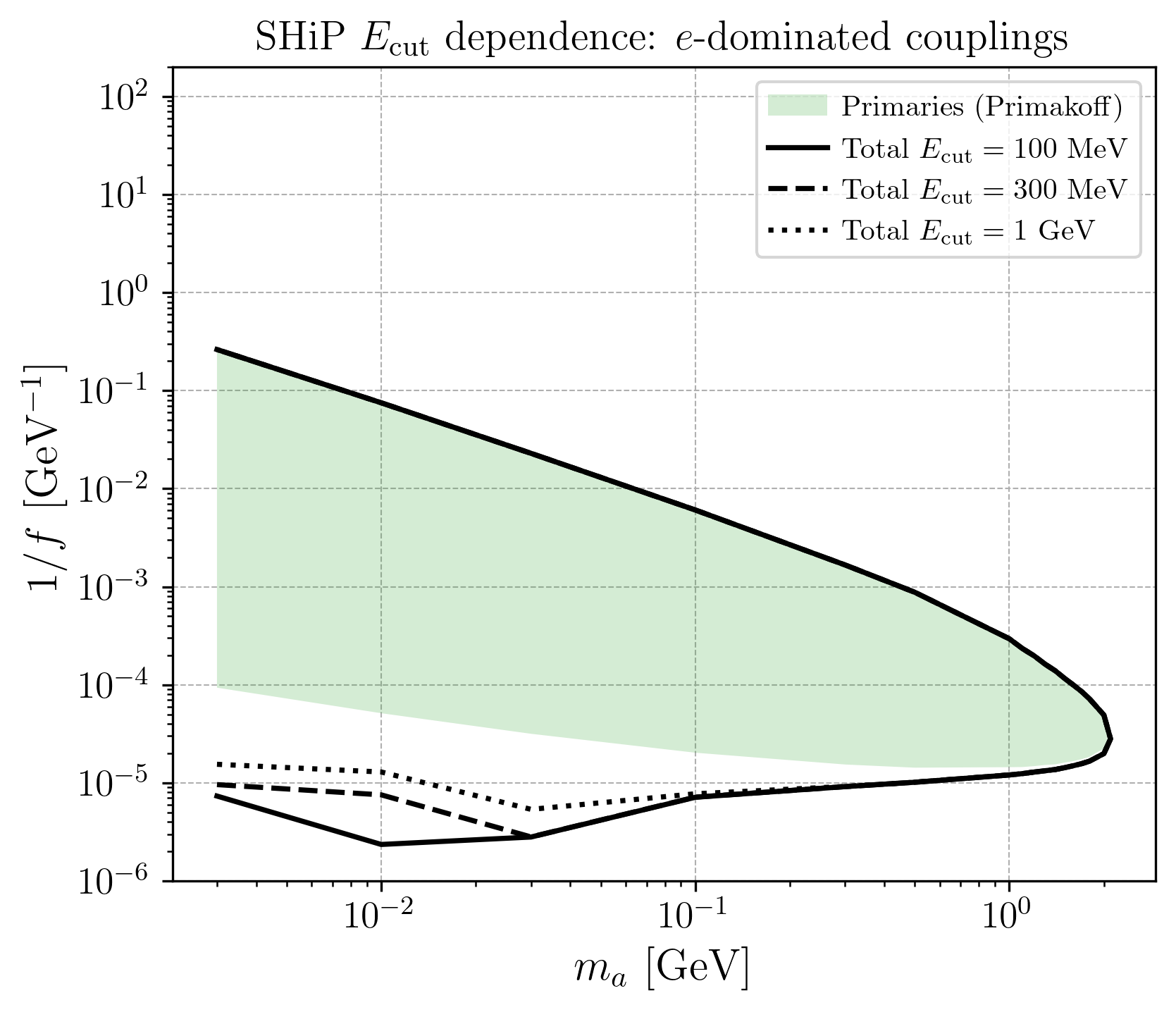}
    \end{minipage}
    \begin{minipage}[t]{0.48\textwidth}
    \centering
    \includegraphics[width=\linewidth]
    {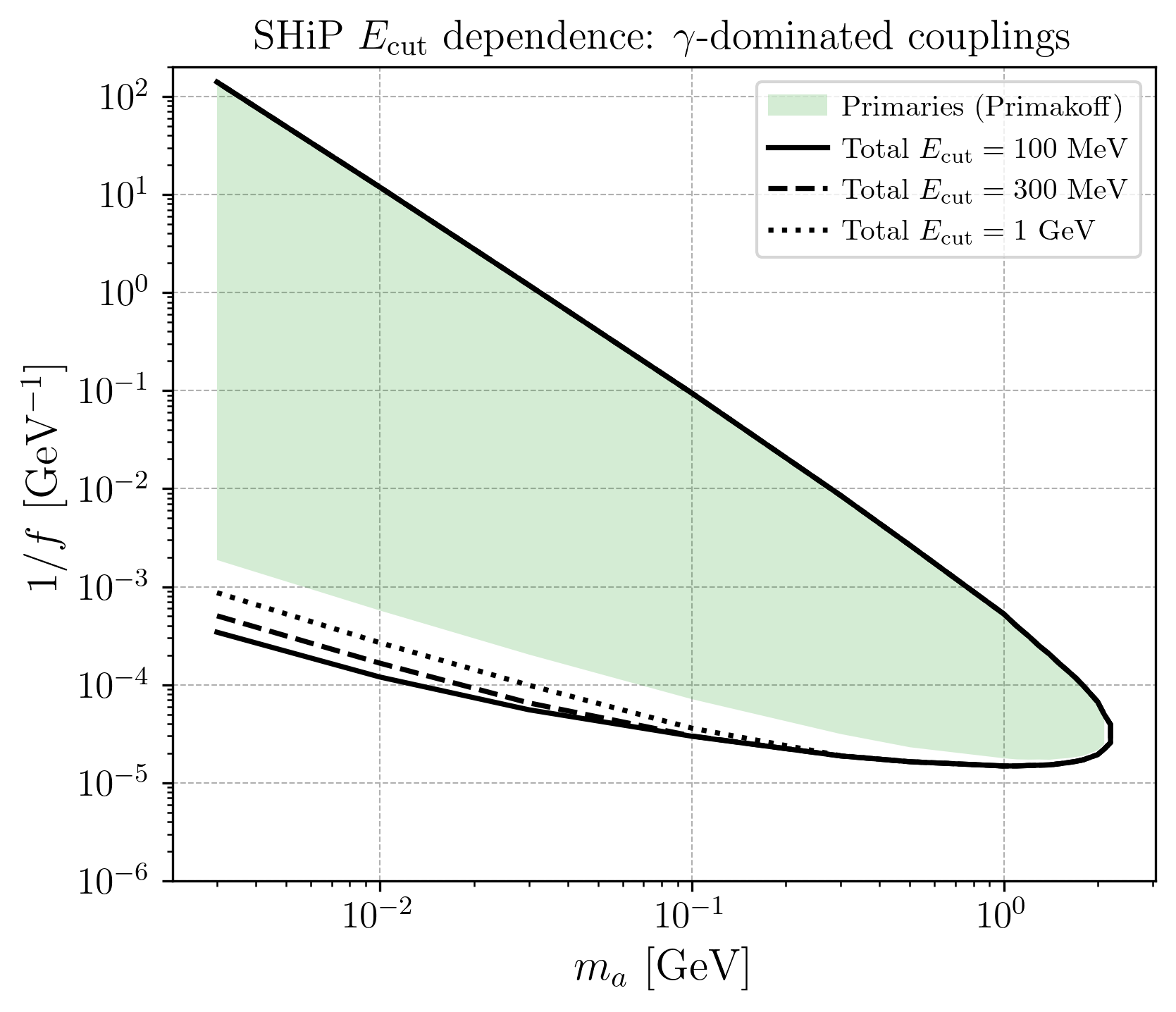}
    \end{minipage}
    \caption{Sensitivity curves of SHiP for 5 signal event in the electron-dominated (\textbf{left panel}) and photon-dominated (\textbf{right panel}) benchmarks with different energy cuts applied. The green shaded region shows the sensitivity for ALPs generated by primary photons arising from neutral meson decays. The solid, dashed, and dotted black line, are indicating the full parameter space covered including the shower contributions for $E_{\text{cut}}=100, \;300,\; 1000$ MeV respectively. We observe (almost) no dependence on the energy cut for the primary reach.}
    \label{fig:ship_ecuts_sens}  
\end{figure*}

\section{Energy thresholds  at SHiP}
\label{app:ecuts}
\begin{figure*}[t]
    \centering
\includegraphics[width=0.32\linewidth]{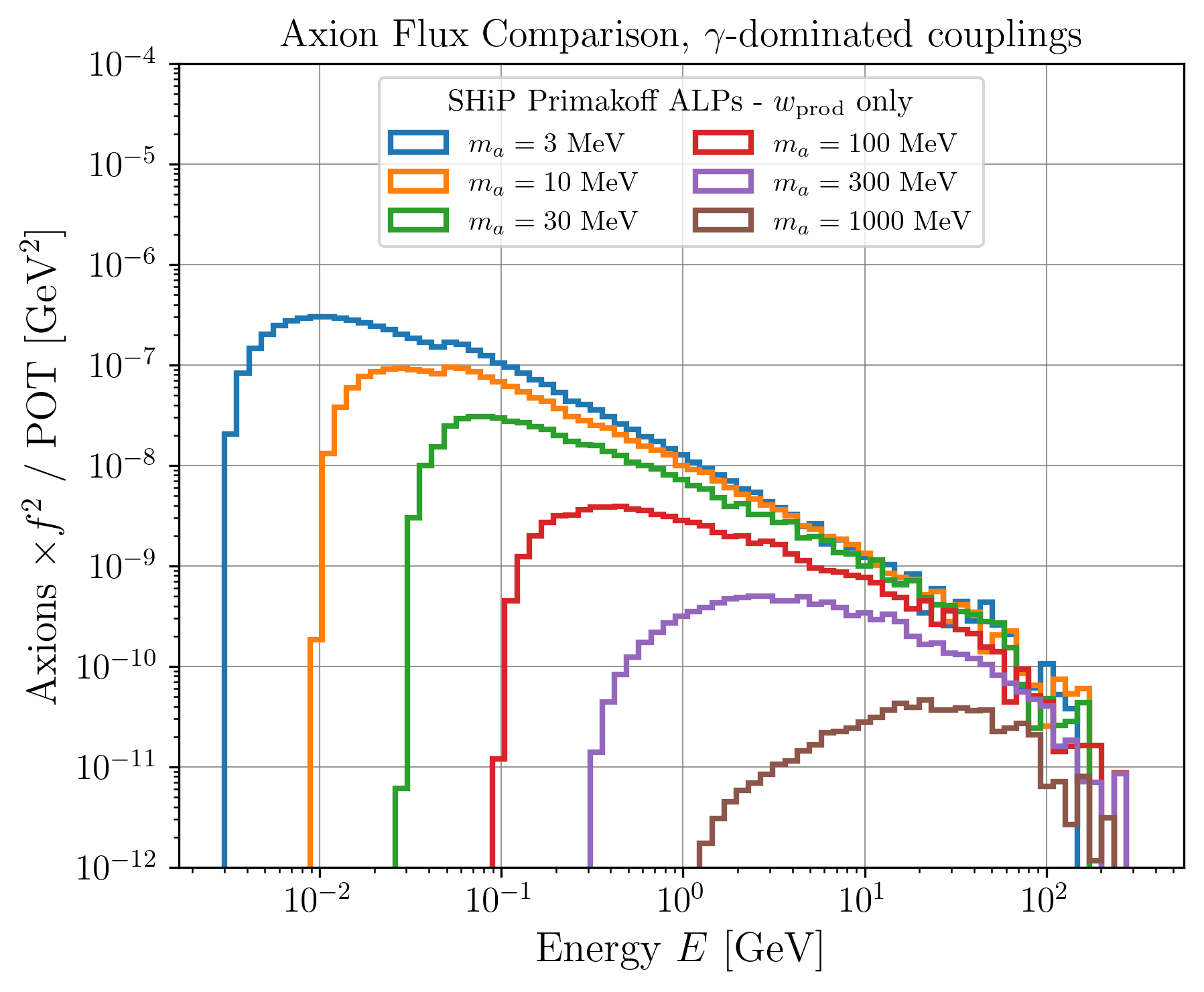}
\includegraphics[width=0.32\linewidth]{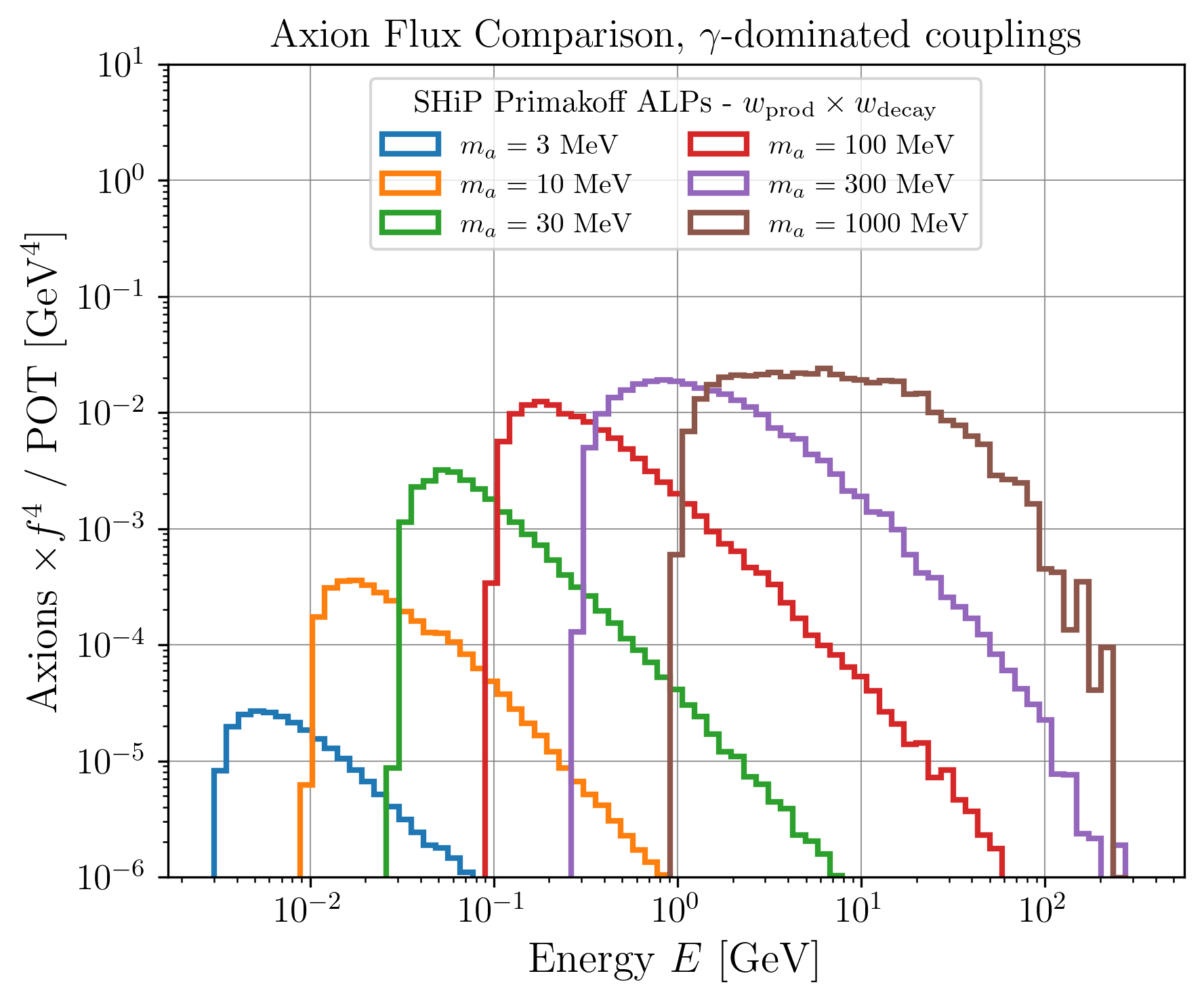}
\includegraphics[width=0.32\linewidth]{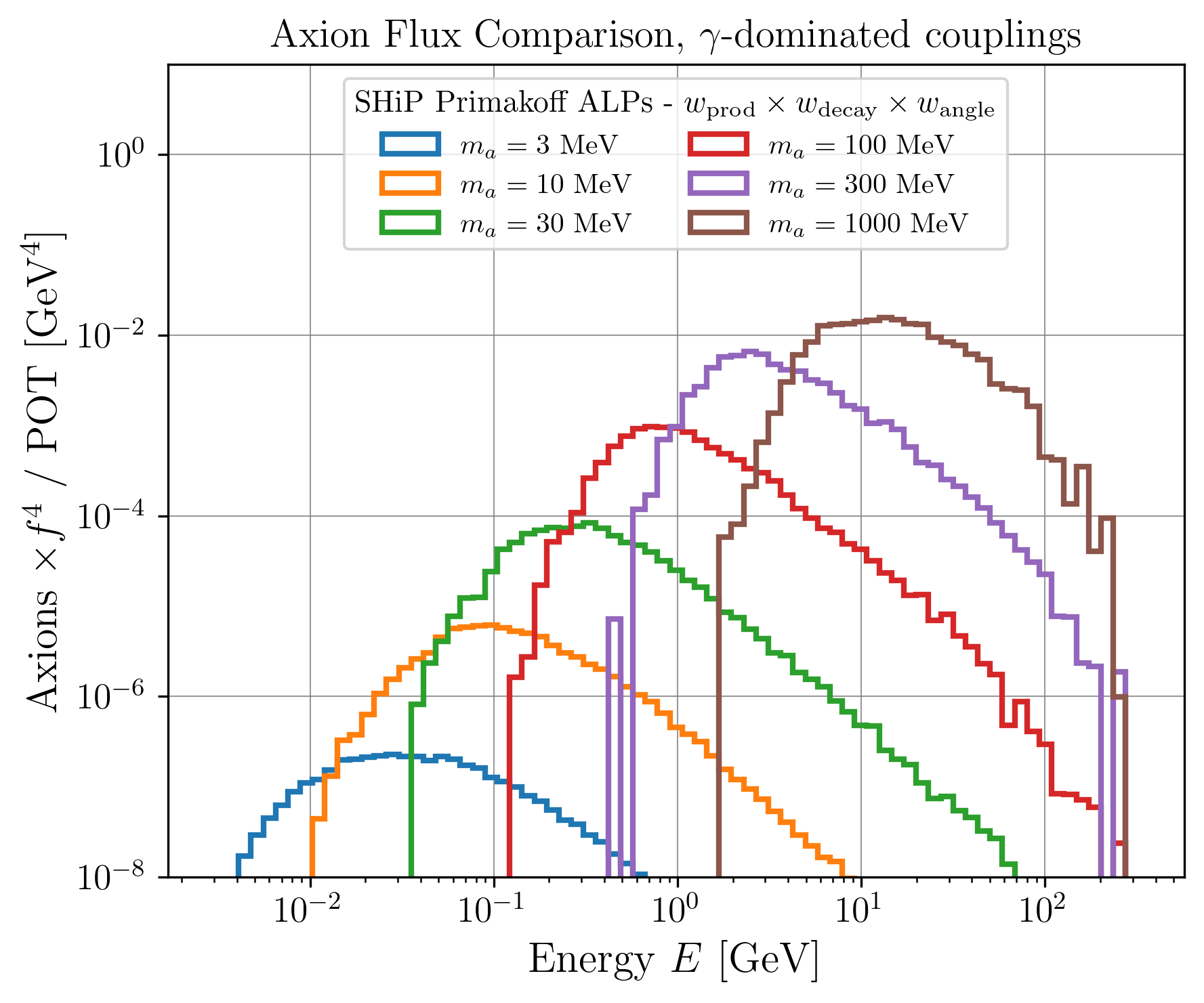}
    \caption{Evolution of the ALP flux through successive application of weights for $m_a = \{3, 10, 30, 100, 300, 1000\}$ MeV; note that we work in the LLP limit, and scale out $1/f^4$ such that weights can be greater than unity. \textbf{Left panel:} Raw production flux, i.e.~with $w_{\text{prod}}$ only, from Primakoff conversion (from both primaries photons arising from neutral meson decay and secondary photons arising from electromagnetic showers) showing enhanced production for lighter ALPs due to the abundance of soft photons in the showers. \textbf{Middle panel:} Decay-weighted flux in the LLP limit ($w_{\text{prod}} \times w_{\text{decay}}$), with high-energy suppression from ALPs escaping the decay pipe; note that $w_{\text{decay}}$ is greater than unity because $1/f$ has been factored out. \textbf{Right panel:} Downstream predicted flux including the angular acceptance of the detector ($w_{\text{prod}} \times w_{\text{decay}}\times w_{\text{geom}} $), with no energy cut applied. The interplay between geometric acceptance and decay kinematics creates a characteristic double-cutoff structure that shifts to higher energies for heavier ALPs. All fluxes are normalized per proton on target (POT) and made coupling-independent by rescaling by the appropriate powers of $f$.}
\label{fig:ship_ecuts_fluxes}
\end{figure*}

In Figs.~\ref{fig:ship_ecuts_sens} and \ref{fig:ship_ecuts_fluxes}, we analyze the impact of changing the energy threshold acceptance at SHiP as well as the effect of the geometric acceptance on the observed flux. As discussed in \cite{Zhou:2024aeu}, the SHiP collaboration can optimize sensitivity to cascade produced long-lived particles by modifying their analysis strategy. In particular lower-energy thresholds are important to profit from the resonant annihilation of high energy positrons with atomic electrons.

In \cref{fig:ship_ecuts_sens}, we show different sensitivity curves at SHiP for different energy cuts, in the electron-dominated scenario (left panel) and photon-dominated one (right panel).
In the electron-dominated benchmark,
the main effect is to impact the resonant annihilation sensitivity, which is one of the dominant active channels at lower masses. 
Indeed, the peak of the resonant flux lies at 
\begin{equation}
E_{\rm res} \simeq \frac{m_a^2}{2 m_e}\simeq100~\text{MeV}\left(\frac{m_a}{10~\text{MeV}}\right)^2.
\end{equation}
For energy cuts $E_{\text{cut}}\le 300~\text{MeV}$, the sensitivity lower bounds for $m_a\ge30~\text{MeV}$ are not affected, and for lower masses, the shower gain at small couplings is still of $\mathcal{O}(10)$, even for thresholds as large as $E_{\text{cut}}\sim 1~\text{GeV}$.
In the photon-dominated scenario, Primakoff is the leading production channel. The effect of the different energy cuts is to suppress the flux (and therefore the gain in sensitivity) for smaller masses.

\cref{fig:ship_ecuts_fluxes} illustrates the progressive application of experimental selections to the ALPs flux in the LLP limit produced via Primakoff conversion for both primary photons and electromagnetic showers. The first panel shows the raw production spectrum (rescaled by $f^{2}$) for various ALP masses $m_a = 3-1000$ MeV, exhibiting a pronounced enhancement at low masses due to the abundance of soft photons in the shower spectrum that efficiently convert when $m_a \ll E_\gamma$. The second panel includes the decay probability weight in the LLP limit $w_{\text{decay}} \approx L/\lambda_i \propto 1/f^2$, which suppresses the high-energy tail where boosted ALPs escape the fiducial volume before decaying.
The third panel incorporates the geometric acceptance weight $w_{\text{geom}}$, introducing a sharp low-energy cutoff at $E_a \sim m_a/\theta_{\text{max}}$ where $\theta_{\text{max}}$ is the angular acceptance of the decay volume. As ALPs' three-momentum goes to zero, they become increasingly transverse and fail the angular selection. The resulting peaked distribution represents the kinematically accessible phase space bounded by geometric acceptance at low energies and decay length considerations at high energies, with the peak position shifting to higher energies for heavier ALPs as expected. 
For SHiP geometry, $\theta_{\text{max}}\sim 0.1$ which implies that $E_{\rm peak}\sim 10\, m_a$. Therefore, in order to fully profit from the shower enhancement for a fixed axion mass $m_a$, the optimal cut for maintaining the signal is $E_{\text{cut}} \lesssim 10 \,m_a$. A detailed study of signal vs.~background is required to understand the optimal cuts for a full experimental analysis. 

\FloatBarrier
\section{$N_{\rm sig}$ dependence}
\label{app:nsig}

\begin{figure*}[t]
  \centering
  \begin{minipage}[t]{0.48\textwidth}
    \centering
    \includegraphics[width=\linewidth]{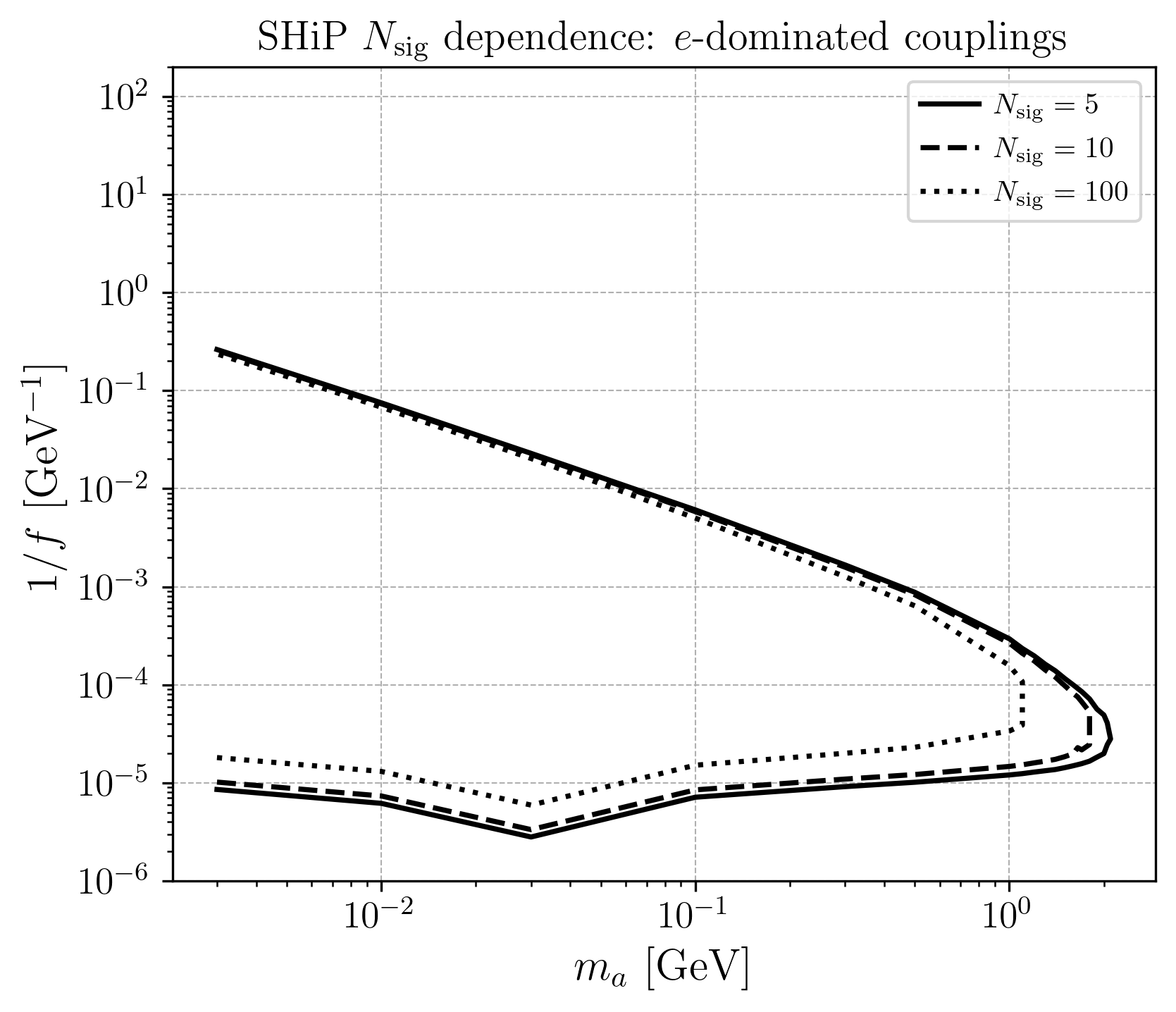}
  \end{minipage}
  \hfill
  \begin{minipage}[t]{0.48\textwidth}
    \centering
    \includegraphics[width=\linewidth]{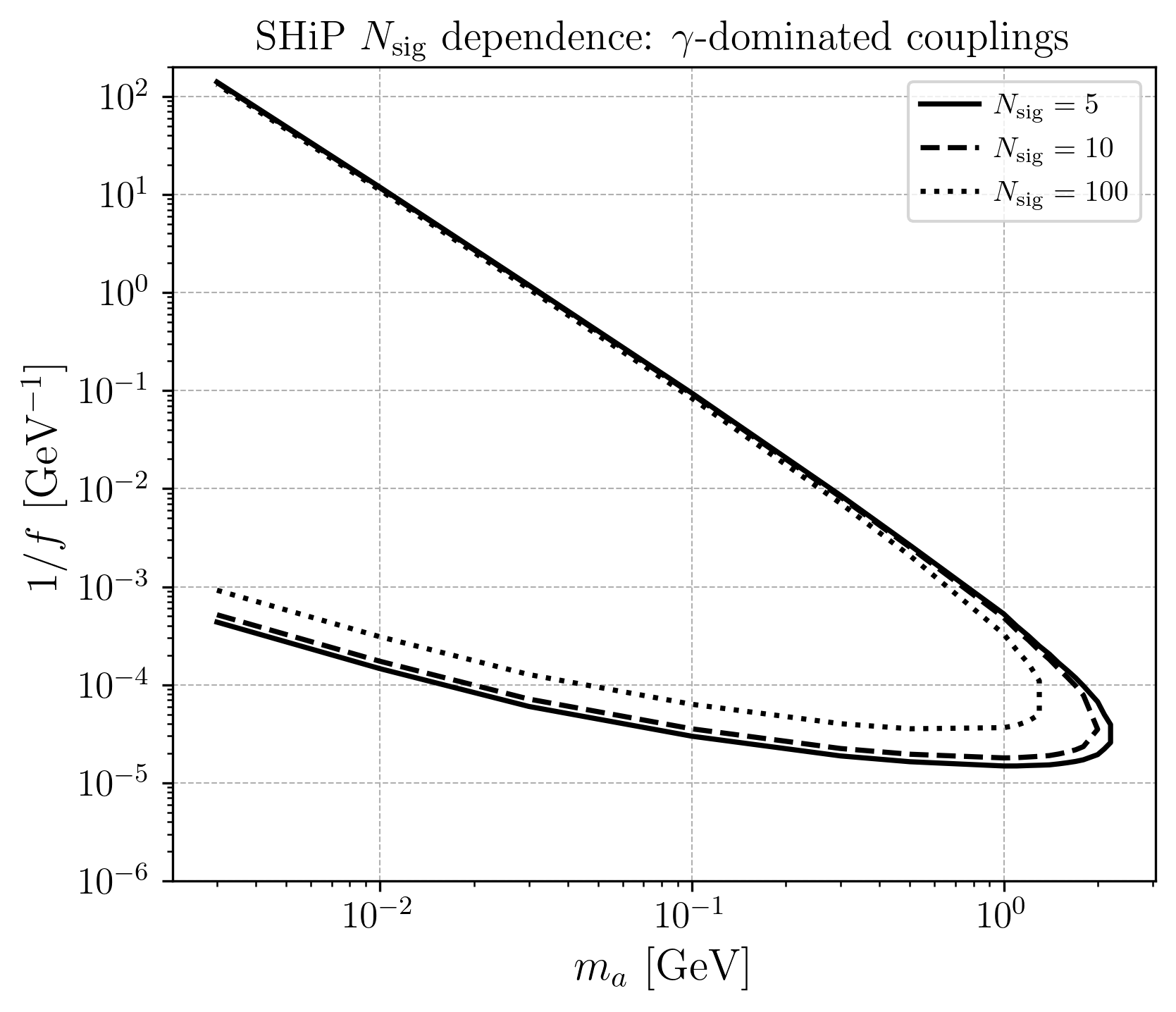}
  \end{minipage}
  \caption{$N_{\rm sig}$ dependence of the projected SHiP sensitivity for the electron-dominated ({\bf left panel}) and photon-dominated ({\bf right panel})
  ALP coupling scenarios. We show contours corresponding to $N_{\rm sig}=5$ (solid), $N_{\rm sig}=10$ (dashed), and $N_{\rm sig}=100$
  (dotted) signal events, assuming an integrated $6\times 10^{20}$ POT and an energy acceptance threshold
  $E_{\text{cut}}=200~\mathrm{MeV}$.}
  \label{fig:ship_nsig}
\end{figure*}

\begin{figure*}[t]
  \centering
  \begin{minipage}[t]{0.48\textwidth}
    \centering
    \includegraphics[width=\linewidth]{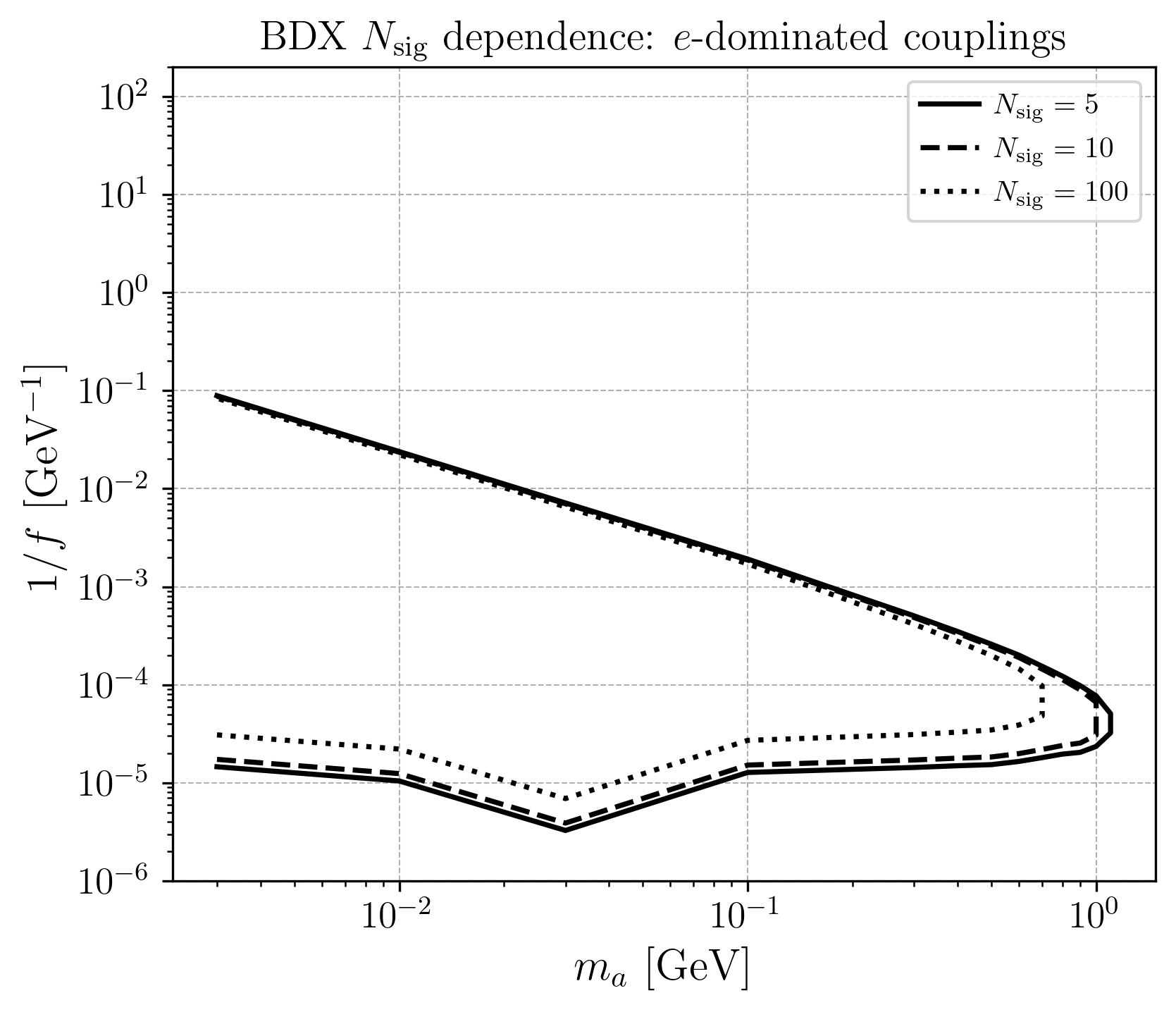}
  \end{minipage}
  \hfill
  \begin{minipage}[t]{0.48\textwidth}
    \centering
    \includegraphics[width=\linewidth]{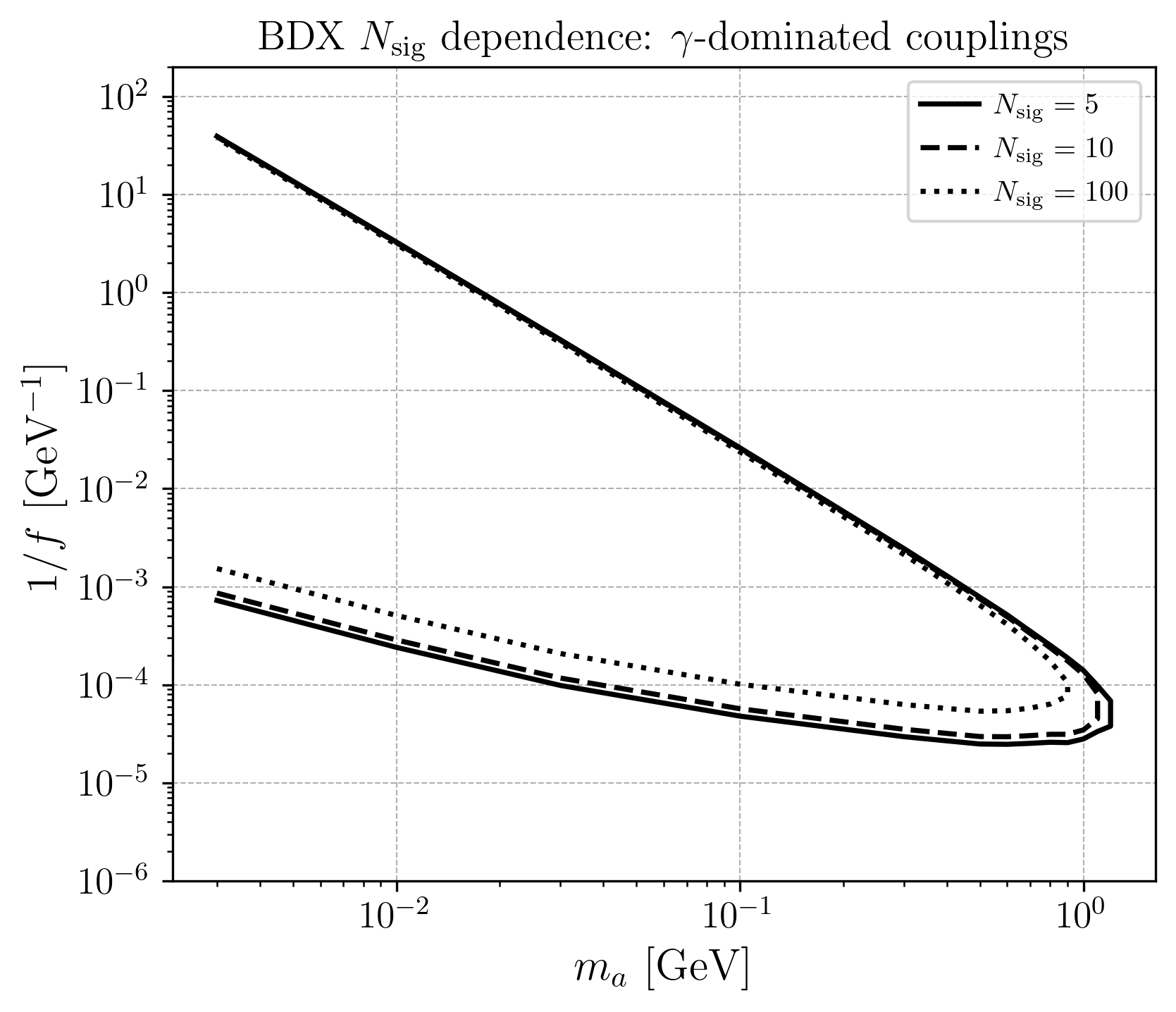}
  \end{minipage}
  \caption{$N_{\rm sig}$ dependence of the projected BDX sensitivity for the electron-dominated ({\bf left panel}) and photon-dominated ({\bf right panel})
  ALP coupling scenarios. We show contours corresponding to $N_{\rm sig}=5$ (solid), $N_{\rm sig}=10$ (dashed), and $N_{\rm sig}=100$
  (dotted) signal events, assuming an integrated $10^{22}$ EOT and an energy acceptance threshold
  $E_{\text{cut}}=300~\mathrm{MeV}$.}
  \label{fig:bdx_nsig}
\end{figure*}

Throughout the main text we present baseline projections assuming background-free operation and a discovery threshold
of $N_{\rm sig}=5$ signal events. Since this choice is somewhat conventional, in this appendix we show how the projected reach changes as the required signal yield is varied. \cref{fig:ship_nsig,fig:bdx_nsig} display the
projected sensitivity contours for $N_{\rm sig}=5,\,10,$ and $100$ events for SHiP and BDX, respectively, for both ALP coupling scenarios considered. As expected, the large-coupling boundary is only weakly affected by the choice of $N_{\rm sig}$, since it is primarily set by the
requirement that the ALP survive to the detector (lifetime/geometry), while the small-coupling region is rate-limited and shifts
more noticeably as $N_{\rm sig}$ is varied.

\section{Primary photon spectra used to seed the SHiP cascade}
\label{app:ship-photon-spectra}

In the SHiP projection, the electromagnetic cascade is seeded by primary on-shell photons produced
predominantly in neutral-meson decays (e.g.\ $\pi^0\to\gamma\gamma$) in the target. To generate the
cascade efficiently, we draw a random subsample of size $N_{\rm samp}$ from the full generated photon
population of size $N_{\rm parent}$ and assign them Monte Carlo weights (cfr.~\cref{eq:mcweight_photons} and related discussion) such that the subsample is a statistically representative proxy for the primary distribution.

\begin{figure*}[t]
  \centering
  \includegraphics[width=\textwidth]{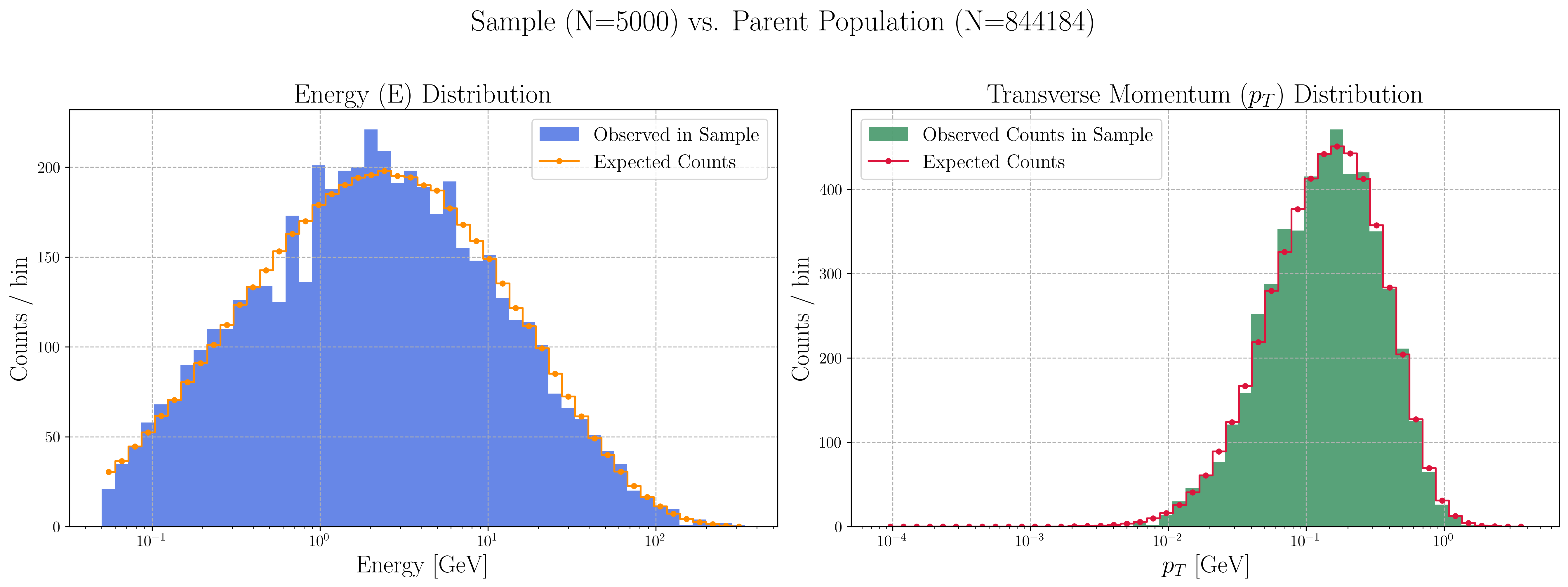}
  \caption{Primary-photon spectra used to seed the SHiP cascade simulation.
  The photon list shown is a random subsample of $N_{\rm samp}=5\times10^3$ photons drawn from a parent population of size $N_{\rm parent}$, which constitute all the photons obtained from the $10^5~pp$ \texttt{PYTHIA}'s collision simulations.
  {\bf Left panel}: photon energy spectrum. {\bf Right panel}: photon transverse-momentum spectrum.
  Histograms show the observed subsample counts, while the curves show the expected counts obtained by scaling the parent-population
  distributions to the subsample size, $N_{\rm samp}/N_{\rm parent}$. Both panels use logarithmic binning on the horizontal axis. Photons with $E\le50~{\rm MeV}$ have been removed from the sampled list.}
  \label{fig:ship-photon-sample-vs-parent}
\end{figure*}

\cref{fig:ship-photon-sample-vs-parent} shows the resulting photon energy and transverse-momentum
spectra for a sampled list used to seed one of the cascade simulations, together with a diagnostic comparison to the
parent population. The histograms show the observed subsample counts, while the curves show the expected
counts obtained by scaling the parent distributions by $N_{\rm samp}/N_{\rm parent}$. The good agreement verifies
that the sampled photon list used as the cascade input reproduces the parent distributions.

\FloatBarrier
\bibliography{mainbib}{}

\end{document}